# Low-Rank Iteration for Efficient Calculations of Topological Surface States

Kaiqi Liu[1], Changsheng He[2], Zihan Xie[1], Shulin Sun[1,3,4], Lei Zhou[2,3,4*], Shaojie Ma[1,3,4*]

[1]College of Future Information Technology, Fudan University, Shanghai 200433, China

[2]State Key Laboratory of Surface Physics and Department of Physics, Fudan University, Shanghai 200433, China

[3]Shanghai Key Laboratory of Metasurfaces for Light Manipulation, Shanghai 200433, China

[4]Key Laboratory of Micro and Nano Photonic Structures (MOE), Fudan University, Shanghai 200433, China

[*]Correspondence: phzhou@fudan.edu.cn (L.Z.) and shaojiema@fudan.edu.cn (S.M.)

## Abstract:

Surface states are hallmarks of topological phases, and their efficient computation is crucial for topology characterization. Conventional supercell methods (SCM) approximate semi-infinite systems as finite periodic ones, but suffer from high computational cost, artificial inter-surface coupling, and slow convergence near band edges. Recently, boundary problems for semi-infinite systems have been recast as block-tridiagonal equations and solved by the cyclic reduction method (CRM) and the transfer matrix method (TMM). However, these approaches remain resource-intensive, as fine-mesh discretization produces very large matrices. Recognizing the intrinsic sparsity and low rank of these matrices, we develop a low-rank iterative method (LIM) with stochastic trace estimation that substantially reduces complexity and memory without sacrificing accuracy, while remaining straightforwardly parallelizable. Benchmarks on 2D and 3D topological systems show orders-of-magnitude acceleration over CRM and TMM. LIM extends readily to heterostructures, defective interfaces, and non-Hermitian systems, providing an efficient platform for topological and plasmonic surface phenomena.

## Introduction

Artificially engineered materials, including photonic crystals, phononic crystals, and metamaterials, provide versatile platforms for exploring topological physics[1-10]. As a central feature, topologically protected surface states enable robust wave transport in devices such as low-loss waveguides and unidirectional components[11-13], and their dispersion and propagation determine key device metrics including operational bandwidth and group velocity[14-16]. Efficient, accurate determination of the surface-state spectrum at a single boundary is therefore crucial for both topological physics and device engineering.

A common approach is the supercell method (SCM), which treats a semi-infinite system as a finite periodic slab and solves its eigenvalue problem for the surface-state dispersion. Suppressing coupling between the two artificial surfaces requires a thick supercell, incurring substantial cost, particularly for finely discretized 3D structures. However, the decay length diverges near band edges, so residual coupling persists at any thickness. Moreover, surface states constitute only a small fraction of the computed bands, embedded among numerous bulk states, so identifying them is nontrivial[17]. A more direct route is the surface Green's function of a partially infinite system[18-23], whose trace defines the surface density of states (SDOS) and links surface states to surface-source excitation spectra. For electronic systems, iterative[24-26], recursive[27,28], and transfer-matrix[29-32] methods compute the local SDOS directly on the block-tridiagonal Toeplitz matrices of tight-binding models, without assembling the full infinite matrix. In contrast, photonic and phononic systems generally exhibit more complex geometries and materials, and require finite-element discretization, producing far more degrees of freedom (DOF) than their electronic counterparts[33-36]. Finite-element block-diagonal formulations recently extended the cyclic-reduction method (CRM) and transfer-matrix method (TMM) to such systems[37,38]. Nevertheless, both costs grow as a power law in the DOFs per unit cell, posing a bottleneck for high-dimensional, finely discretized systems. Thus, applying mature electronic schemes directly to photonic and phononic systems is often prohibitive.

Here, recognizing that the discretization matrices are intrinsically sparse and that the inter-cell coupling matrices are low-rank, we propose a low-rank iterative method (LIM) that overcomes the main cost sources of CRM and TMM for complex photonic and phononic systems (see Fig. 1). For a uniformly discretized unit cell with $N$ DOFs, the rank $r$ of the coupling matrix scales roughly as $N^{1/2}$ and $N^{2/3}$ in 2D and 3D, far smaller than $N$. LIM reformulates the surface Green's function as an iteration of a low-rank Riccati equation restricted to a subspace set by $r$, avoiding operations on full $N \times N$ matrices. The SDOS is then evaluated with Hutchinson's stochastic trace estimator, without explicitly forming the dense Green's function. These low-rank features substantially improve efficiency, enabling mildly super-linear scaling with finite-element DOFs. LIM also applies to systems formed by connecting multiple semi-infinite regions. We demonstrate this on bare and coated crystals, heterojunctions, sandwich structures, and 3D Weyl metamaterials, computing the SDOS and, where applicable, iso-frequency surfaces. Across these geometries, LIM resolves single-interface topological response, eliminates the mixing of opposite-surface states inherent to the SCM, and reduces memory and computation time. It extends directly to non-Hermitian systems, higher-order corner states, and inhomogeneous interfaces, thus offering an efficient, robust tool for topological device design.

## Results

### Surface Green's Function and Density of States

In the frequency domain, the source-free Maxwell equations yield an eigenvalue equation for the electric field $\vec{E}$, as does the fluid phononic wave equation for the acoustic pressure $p$. Despite their distinct physical origins, both share a unified generalized-eigenvalue form with $\omega^2$ as the eigenvalue. The Green's functions $\boldsymbol{G}(\omega; \boldsymbol{r}, \boldsymbol{r}')$ of the photonic [Eq. (1)] and phononic [Eq. (2)] systems are the responses of the corresponding wave operators to point-source excitations $\delta(\boldsymbol{r} - \boldsymbol{r}')$[38]:

$$\underbrace{[\nabla \times \mu^{-1}(\boldsymbol{r}) \cdot \nabla \times - \omega^2 \varepsilon(\boldsymbol{r})]}_{\boldsymbol{M}(\omega;\boldsymbol{r})} \boldsymbol{G}(\omega; \boldsymbol{r}, \boldsymbol{r}') = \boldsymbol{I} \delta(\boldsymbol{r} - \boldsymbol{r}') \quad (1)$$

$$\underbrace{[\nabla \cdot \rho^{-1}(\boldsymbol{r}) \cdot \nabla + \omega^2 K^{-1}(\boldsymbol{r})]}_{\boldsymbol{M}(\omega;\boldsymbol{r})} \boldsymbol{G}(\omega;\boldsymbol{r},\boldsymbol{r}') = \delta(\boldsymbol{r}-\boldsymbol{r}') \tag{2}$$

where $\mu$, $\varepsilon$, $\rho$, and $K$ denote the permeability, permittivity, mass density, and bulk modulus at position $\boldsymbol{r}$. Denoting the differential operator and material terms by $\boldsymbol{M}$, both equations share the same form: $\boldsymbol{M} \cdot \boldsymbol{G} = \boldsymbol{I}$. Physically, by introducing a small positive imaginary component $i \cdot \eta$ into $\omega$, the imaginary part of the Green's function $\boldsymbol{G}(\omega; \boldsymbol{r}, \boldsymbol{r}')$ reflects the response intensity to an excitation at $\boldsymbol{r}'$. Topological systems exhibit distinctive boundary properties, for which the SDOS spectral distribution is the key quantity. The SDOS follows from the surface Green's function $\boldsymbol{G_0}(\omega; \boldsymbol{r}, \boldsymbol{r}')$ for a surface excitation:

$$\mathrm{SDOS}(\omega;\boldsymbol{r}) = \frac{2\omega}{\pi} \cdot \mathrm{Im}\left[\mathbf{tr}\left(\boldsymbol{Q}(\boldsymbol{r}) \cdot \lim_{\eta \to 0^+} \boldsymbol{G}_0(\omega + \mathrm{i}\eta; \boldsymbol{r}, \boldsymbol{r}')\right)\right] \tag{3}$$

Here, $\boldsymbol{Q}$ denotes the metric term, which is $\boldsymbol{\varepsilon}(\vec{r})$ for photonic systems and $\mathbf{K}^{-1}(\vec{r})$ for phononic systems.

The topological systems studied are periodic or quasiperiodic normal to the interface and semi-infinite along it, with identical bulk parameters and translationally invariant inter-cell coupling (see Figs. 1a-b). Eqs. (1) and (2) then take the same block-tridiagonal Toeplitz form after finite-element discretization:

$$\underbrace{\begin{pmatrix} \boldsymbol{A} & \boldsymbol{B} & & \\ \boldsymbol{C} & \boldsymbol{A} & \boldsymbol{B} & \\ & \boldsymbol{C} & \boldsymbol{A} & \boldsymbol{B} \\ & & \ddots & \ddots & \ddots \end{pmatrix}}_{\boldsymbol{M}(\omega;\boldsymbol{r})} \begin{pmatrix} \boldsymbol{G}_0 \\ \boldsymbol{G}_1 \\ \boldsymbol{G}_2 \\ \vdots \end{pmatrix} = \begin{pmatrix} \mathbf{I}_0 \\ \mathbf{0} \\ \mathbf{0} \\ \vdots \end{pmatrix} \tag{4}$$

Here, the diagonal blocks $\boldsymbol{A}$ describe intra-cell coupling, and the off-diagonal blocks $\boldsymbol{B}$ and $\boldsymbol{C}$ represent inter-cell coupling. All blocks are of size $N \times N$ for a unit cell discretized with $N$ DOFs and encode the geometry, materials, and periodicity information . The imaginary part of the trace of $\boldsymbol{G_0}$ yields the SDOS for the boundary excitation $[I_0, 0, 0, \ldots]^{\mathrm{T}}$. Note that $I_0$ can be any boundary excitation, not necessarily constant, to target a specific surface mode, e.g., one with a given polarization.

Boundary modes are conventionally obtained by truncating the semi-infinite $\boldsymbol{M}$ to a finite number of cells in the SCM and diagonalizing the resulting finite matrix, which exploits only the sparsity of $\boldsymbol{M}$ and tolerates cell-to-cell differences. Instead, CRM and TMM compute $\boldsymbol{G_0}$ directly for a truly infinite system[37,38]. CRM iteratively evaluates $\boldsymbol{G_0}$ approximately by eliminating odd-layer Green's functions via Gaussian elimination until the effective inter-layer coupling decays exponentially. By contrast, TMM relates adjacent-layer Green's functions via a transfer matrix and, keeping only decaying eigenmodes, reduces the problem to a linear relation between surface and bulk Green's functions that yields $\boldsymbol{G_0}$ exactly. However, both approaches suffer from computational drawbacks that limit their efficiency: CRM destroys sparsity and iterates on dense matrices of order $N$, whereas TMM obtains the transfer matrix in one step but expands the problem to order $2N$, making TMM slower.

## Low-Rank Iterative Method and Stochastic Trace Estimation

As an illustration, we study a bare semi-infinite 2D photonic crystal: a square-lattice gyromagnetic crystal terminated at the top by a perfect electric conductor (PEC) boundary (see Fig. 2a). The gyromagnetic term breaks time-reversal symmetry, yielding a Chern insulator with a nontrivial Chern number that guarantees unidirectional boundary states[39]. The unit cell is discretized as in Fig. 2b. The governing equation is cast into matrix form $\boldsymbol{M}$, with nodal positions identifying the block rows and columns corresponding to $\boldsymbol{A}$, $\boldsymbol{B}$, and $\boldsymbol{C}$ in Eq. (4). Once the FEM preprocessing is complete, LIM is applied to calculate the SDOS.

Mathematically, solving for the Green's function vector $\boldsymbol{G}$ amounts to inverting the infinite block-tridiagonal Toeplitz matrix $\boldsymbol{M}$:

$$\begin{pmatrix} \boldsymbol{G}_0 \\ \hdashline \boldsymbol{G}_1 \\ \boldsymbol{G}_2 \\ \vdots \end{pmatrix} = \boldsymbol{G} = \boldsymbol{M}^{-1} \cdot \boldsymbol{I} = \boldsymbol{Z} \cdot \boldsymbol{I} = \left(\begin{array}{c:ccc} \boldsymbol{Z}_{00} & \boldsymbol{Z}_{01} & \boldsymbol{Z}_{02} & \cdots \\ \hdashline \boldsymbol{Z}_{10} & \boldsymbol{Z}_{11} & \boldsymbol{Z}_{12} & \ddots \\ \boldsymbol{Z}_{20} & \boldsymbol{Z}_{21} & \boldsymbol{Z}_{22} & \ddots \\ \vdots & \ddots & \ddots & \ddots \end{array}\right) \cdot \begin{pmatrix} \boldsymbol{I}_0 \\ \hdashline 0 \\ 0 \\ 0 \end{pmatrix} \tag{5}$$

Here, $\boldsymbol{Z} = \boldsymbol{M}^{-1}$ and $\boldsymbol{G}_0 = \boldsymbol{Z}_{00}$. The SDOS requires neither the full $\boldsymbol{Z}$ matrix nor $\boldsymbol{Z}_{00}$ itself, but only its trace. By translational invariance, the Green's function blocks satisfy $\boldsymbol{G}_{n+1} = \boldsymbol{S} \cdot \boldsymbol{G}_n$, where $\boldsymbol{S}$ is the transfer matrix. With $\rho(\boldsymbol{S})$ the spectral radius of $\boldsymbol{S}$, i.e., the largest eigenvalue modulus[31], the condition $\rho(\boldsymbol{S}) < 1$ holds in the bandgap, thereby guaranteeing decay away from the boundary[38]. Thus, the transfer matrix $\boldsymbol{S}$ is exactly the stabilizing solution of the block Riccati equation:

$$\boldsymbol{B}\boldsymbol{S}^2 + \boldsymbol{A}\boldsymbol{S} + \boldsymbol{C} = 0 \tag{6}$$

Given $\boldsymbol{A} \cdot \boldsymbol{G}_0 + \boldsymbol{B} \cdot \boldsymbol{G}_1 = \boldsymbol{I}_0$, the surface Green's function satisfies:

$$\mathbf{G}_0 = \boldsymbol{Z}_{00} = (\boldsymbol{A} + \boldsymbol{B}\boldsymbol{S})^{-1} \tag{7}$$

Mathematically, the SDOS amounts rigorously to solving the Riccati equation rather than inverting $\boldsymbol{M}$. This formulation is fully equivalent to the TMM solution[29-32,38]. TMM linearizes the quadratic Riccati equation with auxiliary variables, converting it into a linear eigenvalue problem, but this doubles the eigenvalue-matrix dimension and raises the sparse-inversion cost to $\mathcal{O}((2N)^3)$, thereby substantially reducing efficiency.

To improve efficiency, LIM exploits the low-rank and sparse structure of the finite-element blocks, using a low-rank factorization with an iterative scheme to solve the Riccati equation. In finite-element discretization, the coupling matrices $\boldsymbol{B}$ and $\boldsymbol{C}$ between adjacent cells are set by the mesh connectivity, specifically by the interface DOFs marked orange in Fig. 2b. For $n_{\mathrm{BC}}$ shared DOFs, the coupling matrices involve only these DOFs, so their ranks are bounded by $n_{\mathrm{BC}}$: $r_B, r_C \ll n_{BC}$. For a Hermitian system $\boldsymbol{B} \approx \boldsymbol{C}^\dagger$, the deviation arises only from the small imaginary part $i\eta$ added to $\omega$ for SDOS calculations, hence $r_B = r_C = r$. For the structure in Fig. 2b with $N = 166$, the ranks are only $r = 8$, far smaller than $N$. We factorize the coupling matrices through low-rank sparse QR decompositions:

$$\boldsymbol{X} = \boldsymbol{U}_{\boldsymbol{X}} \boldsymbol{V}_{\boldsymbol{X}}^\dagger, \quad \boldsymbol{X} \in \{\boldsymbol{B}, \boldsymbol{C}\} \tag{8}$$

where $\boldsymbol{U_X}, \boldsymbol{V_X} \in \mathbb{C}^{N\times r_X}$, with $r_X \ll N$. The columns of $\boldsymbol{U_X}$ form orthonormal bases satisfying $\boldsymbol{U_X^*}\boldsymbol{U_X} = \boldsymbol{I}$, and $\boldsymbol{V_X}$ satisfies $\boldsymbol{V_X} = \boldsymbol{X}^\dagger \boldsymbol{U_X}$ (see Fig. 1c).

Since the stable solution $\boldsymbol{S}$ of the Riccati equation describes the exponentially decaying transfer driven by $\boldsymbol{C}$, its column space lies in the right singular subspace of $\boldsymbol{C}$. Physically, $\boldsymbol{S}$ is governed by only a few interface-related modes, reducing $\boldsymbol{S}$ to a lower-dimensional auxiliary factor $\tilde{\boldsymbol{L}}$: $\boldsymbol{S} = \boldsymbol{L}\boldsymbol{V_C^\dagger} \in \mathbb{C}^{N\times N}$, where $\boldsymbol{L} \in \mathbb{C}^{N\times r_C}$ contains the interface-related components and $\tilde{\boldsymbol{L}} = \boldsymbol{V_B^\dagger}\boldsymbol{L} \in \mathbb{C}^{r_B\times r_C}$ is the projected auxiliary factor. Accordingly, $\tilde{\boldsymbol{L}}$ satisfies the self-consistent equation:

$$\tilde{\boldsymbol{L}} = \boldsymbol{\mathcal{F}}(\tilde{\boldsymbol{L}}) = -(\boldsymbol{V}_B^\dagger \boldsymbol{A}^{-1}\boldsymbol{U}_C) + (\boldsymbol{V}_B^\dagger \boldsymbol{A}^{-1}\boldsymbol{U}_B)\cdot\left[\boldsymbol{I} + \tilde{\boldsymbol{L}}\cdot(\boldsymbol{V}_C^\dagger \boldsymbol{A}^{-1}\boldsymbol{U}_B)\right]^{-1}\cdot\tilde{\boldsymbol{L}}\cdot(\boldsymbol{V}_C^\dagger \boldsymbol{A}^{-1}\boldsymbol{U}_C) \quad (9)$$

All SDOS and field-distribution calculations depend only on the low-rank factor $\tilde{\boldsymbol{L}}$ (derivation in Supplementary Section I). Setting $\tilde{\boldsymbol{L}}^0 = 0$, the self-consistent equation is solved by the iteration:

$$\tilde{\boldsymbol{L}}^{k+1} = \boldsymbol{\mathcal{F}}(\tilde{\boldsymbol{L}}^k) = -\boldsymbol{V}_{BAC} + \boldsymbol{V}_{BAB}\cdot\left(\boldsymbol{I} + \tilde{\boldsymbol{L}}^k\cdot \boldsymbol{V}_{CAB}\right)^{-1}\cdot\tilde{\boldsymbol{L}}^k\cdot\boldsymbol{V}_{CAC} \quad (10)$$

where $\boldsymbol{A}_{\boldsymbol{U_X}} = \boldsymbol{A}^{-1}\boldsymbol{U_X} \in \mathbb{C}^{N\times r_X}$ and $\boldsymbol{V}_{\tilde{\boldsymbol{X}}\boldsymbol{A}\boldsymbol{X}} = \boldsymbol{V}_{\tilde{\boldsymbol{X}}}^\dagger \boldsymbol{A}_{\boldsymbol{U_X}} \in \mathbb{C}^{r_{\tilde{X}}\times r_X}$ are constant matrices that can be precomputed. For a topological system, the iteration converges to a unique stabilizing solution and terminates when $\|\tilde{\boldsymbol{L}}^{k+1} - \tilde{\boldsymbol{L}}^k\|/\|\tilde{\boldsymbol{L}}^k\| < \tau$ for a prescribed tolerance $\tau$. During the iteration, all operators are restricted to subspaces of dimension $r_B$ and $r_C$, avoiding explicit manipulation of $N \times N$ dense matrices, which is the key to the mildly super-linear scaling of LIM.

Once $\tilde{\boldsymbol{L}}$ is determined, the surface Green's function $\boldsymbol{G}_0$ can be explicitly derived from Eq. (7) using the Sherman–Morrison–Woodbury formula[40,41]:

$$\mathbf{G}_0 = \mathcal{G}(\tilde{\boldsymbol{L}}) = \boldsymbol{A}^{-1} - \boldsymbol{A}_{U_B}\cdot\left[\boldsymbol{I} + \tilde{\boldsymbol{L}}\cdot\boldsymbol{V}_{CAB}\right]^{-1}\cdot\tilde{\boldsymbol{L}}\cdot\boldsymbol{V_C}^\dagger\cdot\boldsymbol{A}^{-1} \quad (11)$$

Compared with the single-layer solution $\boldsymbol{A}^{-1}$, the second term acts as a correction accounting for the semi-infinite extension. Left-multiplying by $\boldsymbol{Q}$ and taking the imaginary part of its trace directly yields the SDOS. The poles are fully governed by $\boldsymbol{G}_0$, while $\boldsymbol{Q}$ acts only as a location-dependent weight and does not affect dispersion spectra. Thus, $\boldsymbol{Q}$ can be neglected when identifying boundary states, which is convenient when $\boldsymbol{Q}$ is a local matrix difficult to extract. A comparison with and without $\boldsymbol{Q}$ is given in Section II of the Supplementary Material.

A remaining challenge is computing $\boldsymbol{G}_0$ exactly, which requires inverting $\boldsymbol{A}$. Although most matrices are already available from preceding iterations, $\boldsymbol{A}$ is an $N \times N$ sparse matrix, and explicitly forming its inverse generally destroys sparsity, with worst-case time and memory complexities of $O(N^3)$ and $O(N^2)$. Explicit inversion is therefore prohibitive for large sparse problems. To improve efficiency further, we adopt a Hutchinson stochastic trace estimator[42,43] for fast approximation of the SDOS:

$$\mathrm{tr}(\boldsymbol{G}_0) = \mathbb{E}[\boldsymbol{x}^\dagger \boldsymbol{G}_0 \boldsymbol{x}] = \frac{1}{m} \sum_{i=1 \to m} \boldsymbol{x}_i^\dagger \boldsymbol{G}_0 \boldsymbol{x}_i \tag{12}$$

where $\boldsymbol{x}_i$ are independent Rademacher random vectors satisfying $\mathbb{E}[\boldsymbol{x}_i \boldsymbol{x}_i^\dagger] = \boldsymbol{I}$. $\mathrm{tr}(\boldsymbol{G}_0)$ is approximated by averaging quadratic forms over $m$ random vectors ($m \ll N$), instead of all $N$ diagonal entries, with statistical error decreasing as $1/\sqrt{m}$. This requires only the action of $\boldsymbol{A}^{-1}$ on $m$ random vectors: the product $\boldsymbol{A}^{-1} x_i$ is obtained by solving $\boldsymbol{A}\boldsymbol{w}_i = \boldsymbol{x}_i$ via sparse LU back-substitution. Because $\boldsymbol{A}$ is sparse, this back-substitution scales with the sparsity of its LU factors. The same mechanism computes $\boldsymbol{A}^{-1}\boldsymbol{U}_X$. Since both $m$ and $r_X$ are much lower than $N$, these operations are far cheaper than forming $\boldsymbol{A}^{-1}$ explicitly. With typically $m = 50$ evaluations, the SDOS spectral lines are reproduced with high fidelity.

We validate LIM against SCM, CRM, and TMM, as shown in Fig. 2. For SCM, the spectrum of a semi-infinite photonic crystal is approximated by diagonalizing a 9-cell supercell with PEC boundaries at both ends. Surface states protected by the nontrivial Chern number emerge and localize at both boundaries, with the corresponding bands highlighted in orange (upper) and blue (lower) in Fig. 2c. Meanwhile, Figs. 2d–f show the SDOS of an ideal semi-infinite system from CRM, TMM, and LIM, respectively, with LIM agreeing with both while being faster. The SDOS renders bulk states continuous but surface states discrete. A forward-propagating surface state crosses the gap, while the backward-propagating modes coexisting in SCM vanish, underscoring the advantage of the ideal single interface.

Beyond the spectrum, LIM also yields field information that maps onto experimentally accessible boundary excitation fields. Unlike the spectrum, which requires only the trace of $\boldsymbol{G}_0$, the field calculation requires the full $\boldsymbol{G}_0$ and the transfer matrix $\boldsymbol{S}$. Both are explicit in terms of $\tilde{\boldsymbol{L}}$. Once $\boldsymbol{G}_0$ is available, all $\boldsymbol{G}_n$ propagate layer by layer via $\boldsymbol{G}_{n+1} = \boldsymbol{S} \cdot \boldsymbol{G}_n$, yielding the boundary-excited fields. Fig. 2g compares the SCM eigenstate with the LIM Green's function, showing excellent agreement and confirming field localization near the boundary.

**Numerical Examples for Various Topological Systems**

LIM is not restricted to 2D bare systems. For systems with cladding layers, heterojunctions, sandwich structures, or more complex 3D geometries, the mathematical structure is similar. The low-rank iteration solves the Riccati equation in each semi-infinite region in the same way. The interface coupling gives expressions for $\boldsymbol{G}_0$ analogous to Eq. (7), and the stochastic trace estimator then yields the SDOS efficiently. We further demonstrate LIM with five representative examples: Fig. 3 summarizes the dispersions of four distinct 2D topological systems, and Fig. 4 presents a 3D Weyl system with its dispersion and iso-frequency surfaces. The formulation procedure and pseudocode are in Section III of the Supplementary Material.

The first 2D example is a coated Chern insulator, identical to Fig. 2 except that a cladding layer is inserted between the upper interface and the PEC boundary, and every third cell along the boundary is enlarged to fold the Brillouin zone to one third of its size (see Figs. 3a–c). SCM bands for a 9-cell supercell and the LIM SDOS are shown in Figs. 3b-c. Bragg scattering complicates the bands and hinders mode selection in SCM. Nevertheless, the SDOS contains only clean single-interface topological states with positive group velocity, localized near the upper surface and matching the orange bands in Fig. 3b.

The second 2D example is a heterojunction of two semi-infinite valley Chern insulator photonic crystals (see Figs. 3d–f). Opposite rod orientations give the two sides distinct valley Chern numbers, so valley boundary states propagate bidirectionally along the junction. Figs. 3e-f show SCM bands for a 19-cell supercell and the LIM SDOS. Two boundary bands with opposite group velocities appear within the valley gap, in excellent agreement with SCM.

The third 2D example is a hexagonal-lattice photonic crystal with a single defect layer, combining semi-infinite crystals and a sandwich layer (see Figs. 3g–i). Triangular air holes of varying size and orientation are etched in a dielectric slab. Unequal inclusion sizes in the upper and lower crystals break spatial symmetry and open a gap near the Dirac point, whereas equal-size inclusions in the middle layer restore the degeneracy locally, producing localized modes that propagate bidirectionally. SCM bands for a 19-cell supercell and the LIM SDOS are in Figs. 3h-i. The two agree well, and LIM confines the results to the single interface.

The last 2D example is a phononic topological insulator, formed by joining two semi-infinite hexagonal-lattice phononic crystals with different inclusion radii (see Figs. 3j–l). As in the photonic case, once the $\boldsymbol{M}$ matrix is obtained by FEM, the SDOS follows the same procedure. SCM surface bands for a 9-cell supercell and the LIM SDOS are shown in Figs. 3k-l. Two pseudospin modes with opposite group velocities form interface-state bands traversing and connecting the band gap, again in excellent agreement.

The preceding examples are 2D, with scalar fields discretized by Lagrange finite elements with node-based DOFs. The final example extends LIM to the SDOS and constant-frequency surfaces of a 3D semi-infinite Weyl metamaterial (Fig. 4). In 3D problems with curl operators, Nédélec edge elements are used, with low-order DOFs defined by tangential integrals along mesh edges[44]. Despite these differences, LIM still exploits the sublinear rank of the interface-related matrices. The 3D ideal Weyl metamaterial[45] consists of chiral metallic elements, with a PEC-terminated surface. Fig. 4b shows SCM surface bands for a 9-cell supercell, with top and bottom surface states in orange and blue. Figs. 4c-d show the LIM SDOS for the two interfaces, examined separately through the boundary conditions, in excellent agreement with SCM. Fig. 4e shows the SDOS equifrequency contours on the top surface from 10 to 12 GHz: the mode projections of the two oppositely chiral Weyl points are always linked by Fermi-arc surface states, in excellent agreement with Weyl-topology predictions.

**Computing Efficiency and Accuracy**

This section assesses the efficiency and accuracy of LIM, which mainly consists of two steps: solving $\tilde{\boldsymbol{L}}$ by low-rank iteration and evaluating $\mathrm{tr}(\boldsymbol{G}_0)$ by a stochastic trace estimator.

We consider a Hermitian system with low-rank factors satisfying $r_B = r_C = r \ll N$. The Riccati iteration for $\tilde{\boldsymbol{L}}$ runs entirely in the low-rank subspace: each step involves only $r \times r$ matrices and costs $O(kr^3)$ for $k$ iterations. Since $r \ll N$, the main computational cost arises from initialization and applications of $\boldsymbol{A}^{-1}$. The low-rank QR decompositions of $\boldsymbol{B}$ and $\boldsymbol{C}$ cost approximately $O(Nr^2)$, while the $\boldsymbol{A}^{-1}$-related initialization comprises a sparse LU factorization with cost $O(s)$, where $s$ denotes the number of LU nonzeros. With optimal nested dissection, $s$ scale as $O(NlogN)$ in 2D and $O(N^{4/3})$ in 3D[46,47]. Thus, the precomputation of $\boldsymbol{A}^{-1}\boldsymbol{U}_X$ cost $O(rs)$, and the low-rank projections of $\boldsymbol{V}_X^\dagger\boldsymbol{A}^{-1}\boldsymbol{U}_X$ cost $O(Nr^2)$, totaling $O(rs + Nr^2)$ for initialization. The same $\boldsymbol{A}^{-1}$-related operations dominate stochastic trace estimation: each of the $m$ random vectors, typically $m = 50$, requires computing $w_i = \boldsymbol{A}^{-1}x_i$ by sparse LU back-substitution costing $\mathcal{O}(s)$ plus $\mathcal{O}(Nr)$ matrix–vector products. Thus, the overall complexity is thus $\mathcal{O}[rs + Nr^2 + kr^3 + m \cdot (s + Nr)]$. Because $r$-dependent terms have lower power in $N$ and small pre-factors, the cost is dominated by sparse $\boldsymbol{A}^{-1}$-related operations, scaling as $\mathcal{O}[(r + m) \cdot s]$.

For concreteness, we measure the actual cost. For different DOFs, we compute the SDOS of the system in Fig. 2a at 51 frequencies using LIM, LIM-invA, CRM, and TMM, recording peak memory and time. Here, LIM-invA solves $\tilde{\boldsymbol{L}}$ by low-rank iteration and evaluates $\boldsymbol{G}_0$ directly, without stochastic trace estimation. The comparison thus isolates the benefit of stochastic trace estimation over direct $A^{-1}$ computation. The gain has a clear structural origin (see Figs. 5a,b): the boundary DOFs $n_{BC}$, the coupling rank $r_B$, and the nonzeros $\boldsymbol{s}_i$ of $\boldsymbol{A}$ and $\boldsymbol{L}_{\boldsymbol{A}}$ all grow slowly with $N$. Specifically, the $r$-dependent terms $\mathcal{O}(Nr^2)$ and $\mathcal{O}(kr^3)$ grow super-linearly with $N$ as $r \sim \mathcal{O}(N^{0.56})$ in 2D, but their small pre-factors keep sparse $\boldsymbol{A}^{-1}$ operations dominant. The bottleneck is thus the LU factors, which grow as $s \sim \mathcal{O}(N^{1.27})$, only moderately steeper than the $\mathcal{O}(NlogN)$ expected for optimal nested dissection. Consequently, the empirical memory and time complexities are only $\mathcal{O}(N^{1.17})$ and $\mathcal{O}(N^{1.25})$ (see Figs. 5c,d). By contrast, CRM scales as $\mathcal{O}(N^{2.03})$ in memory and $\mathcal{O}(N^{2.48})$ in time, whereas TMM requires $\mathcal{O}(N^{2.03})$ memory and $\mathcal{O}(N^{3.31})$ time. LIM is thus substantially more efficient for large-scale 2D systems.

For 3D systems, the LU factorization suffers far more fill-in, so each application of $\boldsymbol{A}^{-1}$ is more expensive and the gain of LIM is somewhat smaller than in 2D, though still substantial. A rigorous asymptotic analysis is not pursued because CRM and TMM become prohibitively costly. Instead, we benchmark a 3D structure with 4,643 DOFs, averaging over 20 independent runs at different frequencies (Table 1). Compared with CRM, LIM consumes about 1/17 of the memory and 1/15 of the time; relative to TMM, about 1/16 of the memory and only 1/184 of the time. LIM thus retains a clear 3D advantage, reducing both memory and time by more than one order of magnitude.

Finally, we remark on the accuracy of LIM. Indeed, Figs. 2–4 already demonstrate remarkable accuracy despite high efficiency. Once the low-rank iteration converges, accuracy is fully determined by the prescribed tolerance $\tau$ and can be regarded as numerically exact. For stochastic trace estimation, accuracy improves with the number of random vectors $m$: as shown in Fig. 5e, the root-mean-square error (RMSE) decreases as $1/\sqrt{m}$ with increasing $m$. The qualitative SDOS features are usually of primary interest, and $m = 50$ already suffices for representative surface states. Increasing $m$ further reduces the statistical error, approaching explicit evaluation, but the cost grows approximately linearly with $m$. A trade-off between accuracy and efficiency is therefore required.

## Discussion

In summary, LIM efficiently computes the SDOS of semi-infinite topological photonic and phononic systems. Exploiting the low-rank and sparse structure of FEM discretization, it resolves the surface response by low-rank Riccati iteration and evaluates the SDOS by stochastic trace estimation without constructing large-scale matrices. Operating directly on the Green's function, it avoids the boundary-coupling issues of SCM. LIM matches Green's-function-based CRM and TMM in accuracy while substantially reducing memory and time, with advantages growing with DOFs and making it practical for large-scale topological systems.

Beyond photonic and phononic systems, LIM holds promise for electronic-structure calculations with localized bases such as numerical atomic orbitals (NAOs)[48] and maximally localized Wannier functions[49]. For complex structures that are challenging for conventional electronic-structure methods, such as large supercells and quasi-one-dimensional nanoribbons, the Hamiltonian retains the same sparse, low-rank block-tridiagonal form, so low-rank Riccati iteration could compute the embedding self-energies of semi-infinite leads without constructing the full device Hamiltonian, providing an efficient alternative to recursive Green's function methods.

# Methods

## Simulation environment

All simulations reported in this work, both the numerical examples and the performance benchmarks, were carried out with COMSOL Multiphysics, controlled through a MATLAB scripting interface. COMSOL serves here primarily as a mesh generator; other open-source finite-element meshing packages could fulfill the same role. In a typical run, a supercell is constructed in COMSOL, and its finite-sized eigenmatrix $\boldsymbol{M}$ is exported to MATLAB. There, $\boldsymbol{M}$ is split into the diagonal self-coupling blocks $\boldsymbol{A}$ and the off-diagonal inter-coupling blocks $\boldsymbol{B}$ and $\boldsymbol{C}$, which LIM, CRM, and TMM all act on. Since every method is applied to the same geometry and the same mesh, the performance metrics, including computational time, memory usage, and accuracy, are compared under strictly identical conditions, so the comparison is rigorous and free of bias. Timing and memory were measured on a personal computer with an Intel Core i7-13700 CPU (16 cores, 24 threads, 2.10 GHz).

## Details of numerical examples

All structures were discretized with the finite-element method. The specific structural and material settings are provided in the simulation files archived in the open data repository[50].

The bare photonic crystal in Fig. 2 is based on a square lattice with lattice constant $a$, and its unit cell contains a cylindrical inclusion of radius $0.13a$ with relative permittivity $\varepsilon_r = 13$ and relative permeability $\mu_{ii} = 1$, $\mu_{xy} = -\mu_{yx} = 0.4i$. This gyromagnetic response breaks time-reversal symmetry and yields the Chern-insulator phase. The coated crystal in Figs. 3a–c shares this geometry, except that an additional air layer of thickness $d = 0.5a$ separates the upper interface from the PEC boundary. Along this boundary, every third boundary unit cell is slightly enlarged, with its inclusion radius raised to 0.2a, while the rest keep the original $0.13a$. This enlargement folds the Brillouin zone to one third of its original size.

The valley-Chern heterojunction in Figs. 3d–f joins two semi-infinite photonic crystals of triangular dielectric rods with side length $0.615a$, chamfer length $0.11a$, relative permittivity $\varepsilon_r = 1$ and relative permeability $\mu_r = 13$. Opposite rod orientations give the two sides distinct valley Chern numbers. The defect-layer structure in Figs. 3g–i consists of a dielectric slab with relative permittivity $\varepsilon_r = 11.56$ and permeability $\mu_r = 1$, perforated by triangular air holes of side length $1.3a/\sqrt{3}$ and $0.7a/\sqrt{3}$ in the upper and lower crystals, respectively, with equal-size holes in the middle layer. The phononic crystal in Figs. 3j–l joins two hexagonal-lattice acoustic crystals whose cylindrical inclusions have radii $0.3a$ and $0.45a$.

The 3D Weyl metamaterial in Fig. 4 consists of saddle-shaped chiral metallic elements embedded in a PCB substrate with relative permittivity $\varepsilon_r = 2.2$, with in-plane periods $a_x = a_y = 3mm$ and out-of-plane period $a_z = 6mm$. The element geometry is specified by $d_p = 0.75mm$, $d_m = 1.6mm$, $t_m = 0.2mm$ and $R_m = 0.4mm$. Both the top and bottom surfaces are terminated by PEC boundaries.

## Acknowledgements

**Funding**: This work was supported by National Key Research and Development Program of China (Grants No. 2023YFA1407700, 2023YFA1406901, and 2022YFA1404700), the Key Program of the National Natural Science Foundation of China (No. 12634017), National Natural Science Foundation China (Grants No. 12374343, 12221004, 62192771), and the start-up funding of Fudan University (JIH1232098Y). We also thank Dr. Yixin Sha for helpful discussions.

**Author contributions:** S.M. conceived the initial concept, K.L., C.H., Z.X. and S.M. performed numerical simulations and analytical calculations, S.S., L.Z. and S.M. participated in the analysis of the results, and S.M. and L.Z. supervised the project.

**Competing interests:** The authors declare no conflict of interest.

**Data and materials availability:** All data and source code that support the findings of this study are openly available in the Science Data Bank[50] and via GitHub at https://github.com/KaiqiLiu-LKQ/Low-Rank-Iteration-Method-for-Topological-Surface-States.

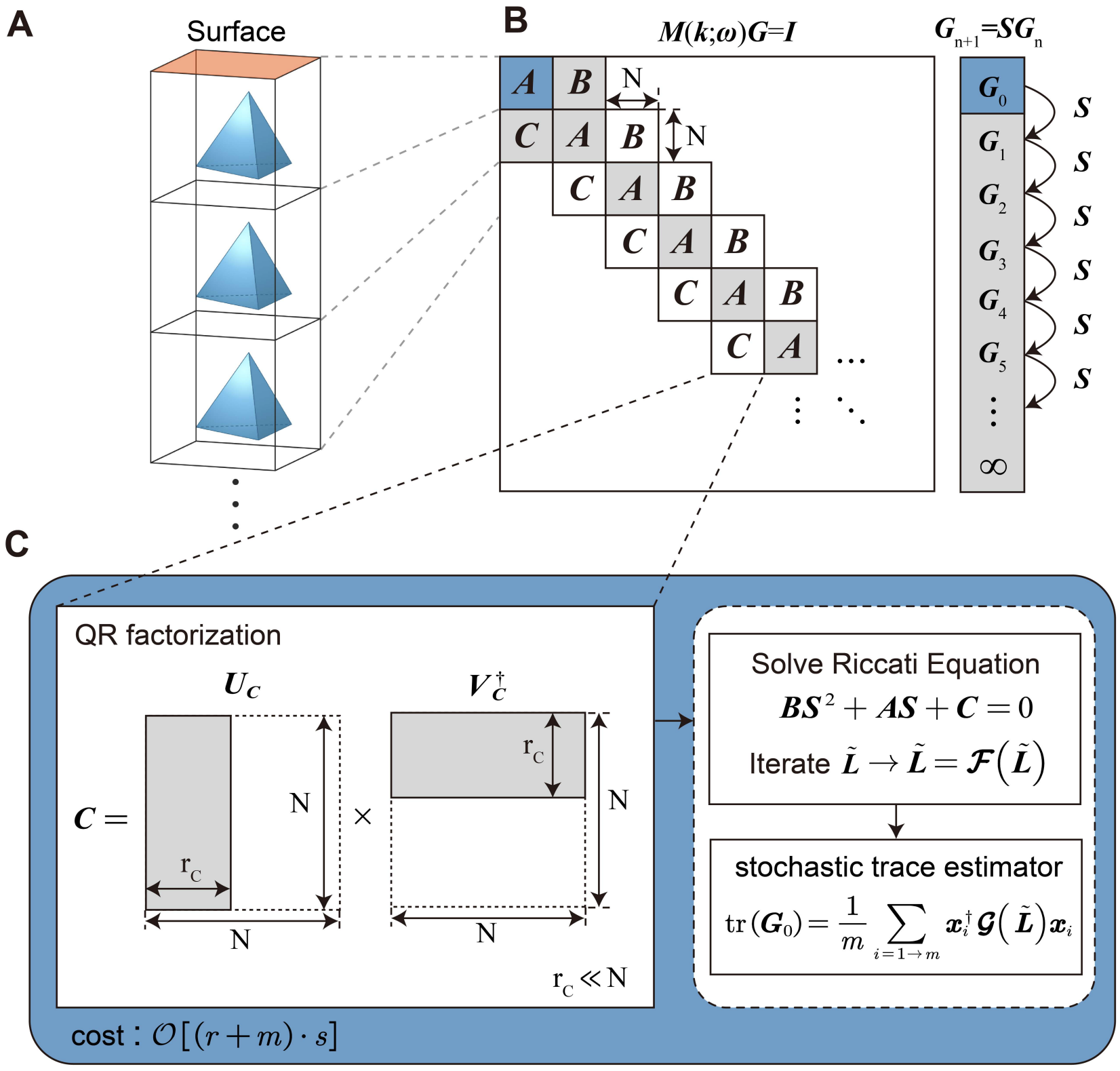

**Fig. 1. The Low-Rank Iterative Method.** (a) Schematic of a representable semi-infinity array structure. (b) Each cell is represented by a sparse block matrix $\boldsymbol{A}$ coupled by low-rank sparse coupling matrices $\boldsymbol{B}$ and $\boldsymbol{C}$ in FEM, with blocks of the Green's function satisfying $\boldsymbol{G}_{n+1} = \boldsymbol{S}\boldsymbol{G}_n$. (c) Schematic of low-rank QR factorization in LIM. The low-rank nature of $\boldsymbol{C}$ ensures that the QR decomposition yields $\boldsymbol{U}_C$ and $\boldsymbol{V}_C$ containing only a finite number $r_C$ of orthonormal bases, thereby reducing the complexity of the iteration.

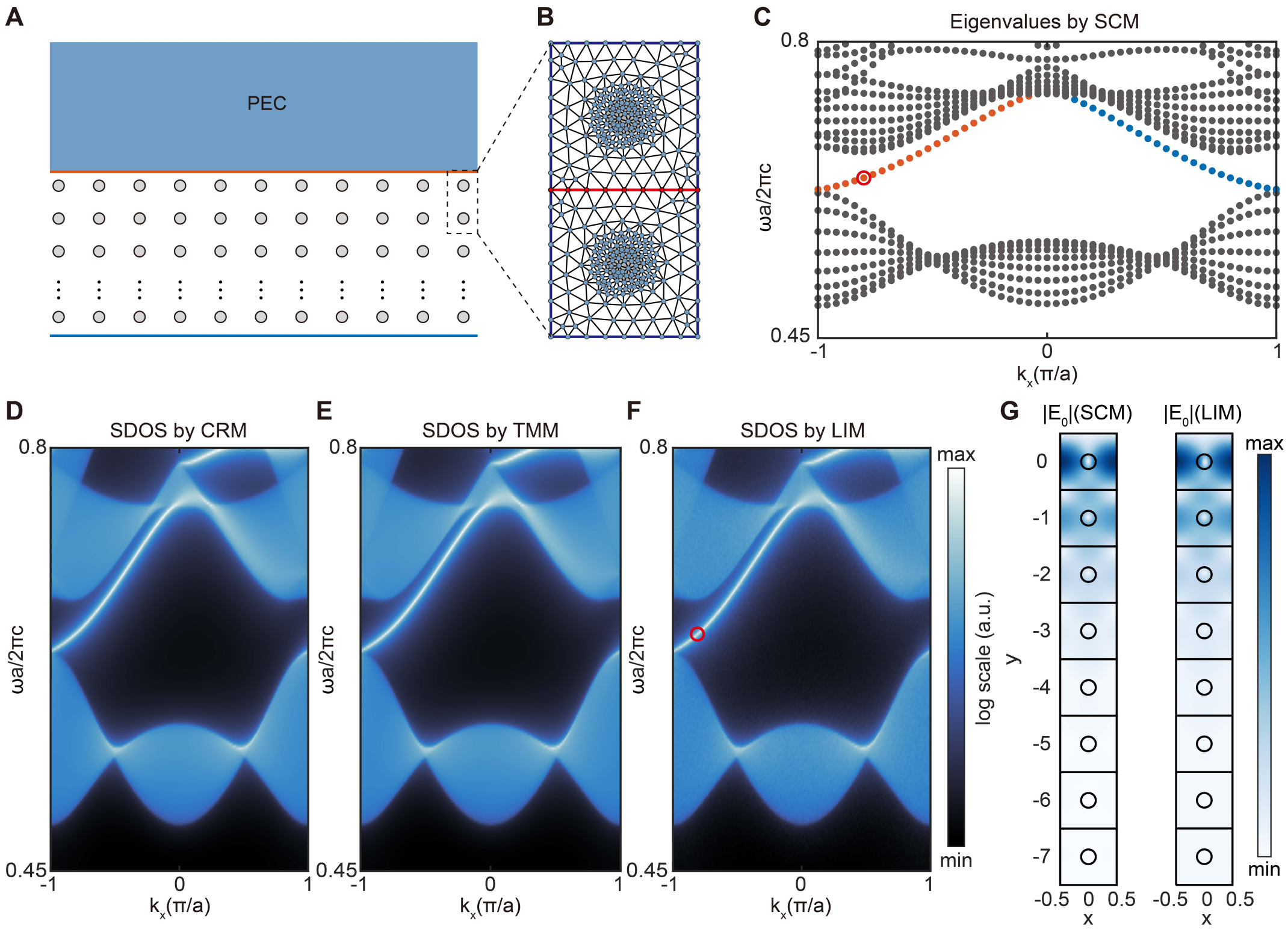

**Fig. 2. SDOS on a Semi-infinite Photonic Crystal.** (a) Schematic of the semi-infinite photonic crystal with a PEC boundary at the top. (b) Discretized unit cell mesh by FEM, where nodes inside the unit cell and those shared between unit cells (highlighted in orange) contribute to the intra-cell and inter-cell terms, corresponding to the block matrix $\boldsymbol{A}$ and the coupling matrices $\boldsymbol{B}$ and $\boldsymbol{C}$ in Fig. 1b, respectively. (c) Band structure of a 9-cell supercell obtained by SCM, with surface states located near the upper and lower PEC boundaries highlighted in orange and blue, respectively. (d-f) SDOS calculated by CRM, TMM, and LIM, respectively. (g) The distribution of the eigenstate obtained by SCM and of the Green's function solved by LIM, respectively.

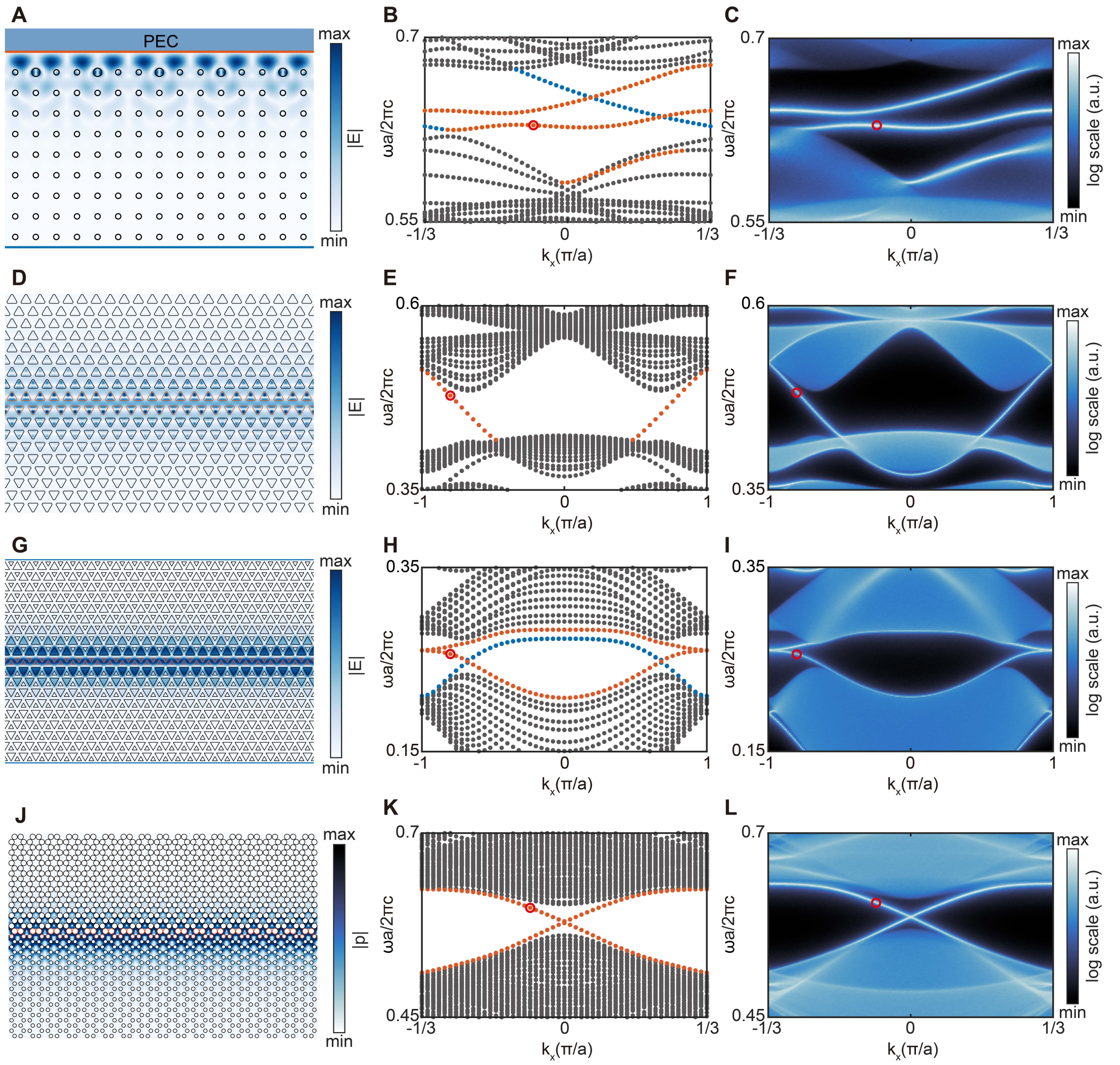

**Fig. 3. Topological Surface States in different systems.** (a-c) Semi-infinite photonic Chern insulator coated with a dielectric plate and terminated by a PEC boundary: (a) schematic, (b) band structure of a supercell obtained by SCM, and (c) SDOS spectrum calculated by LIM, with the representative field distribution shown in (a). (d−f) Heterojunction formed by joining two semi-infinite valley Chern insulator photonic crystals. (g–i) Hexagonal-lattice photonic crystal containing a single defect layer. (j–l) Phononic topological insulator formed by splicing two semi-infinite crystals. Panels (d–l) follow the same layout as (a–c). In the SCM results, modes at the target interface are marked in orange, while those at the other boundaries are marked in blue.

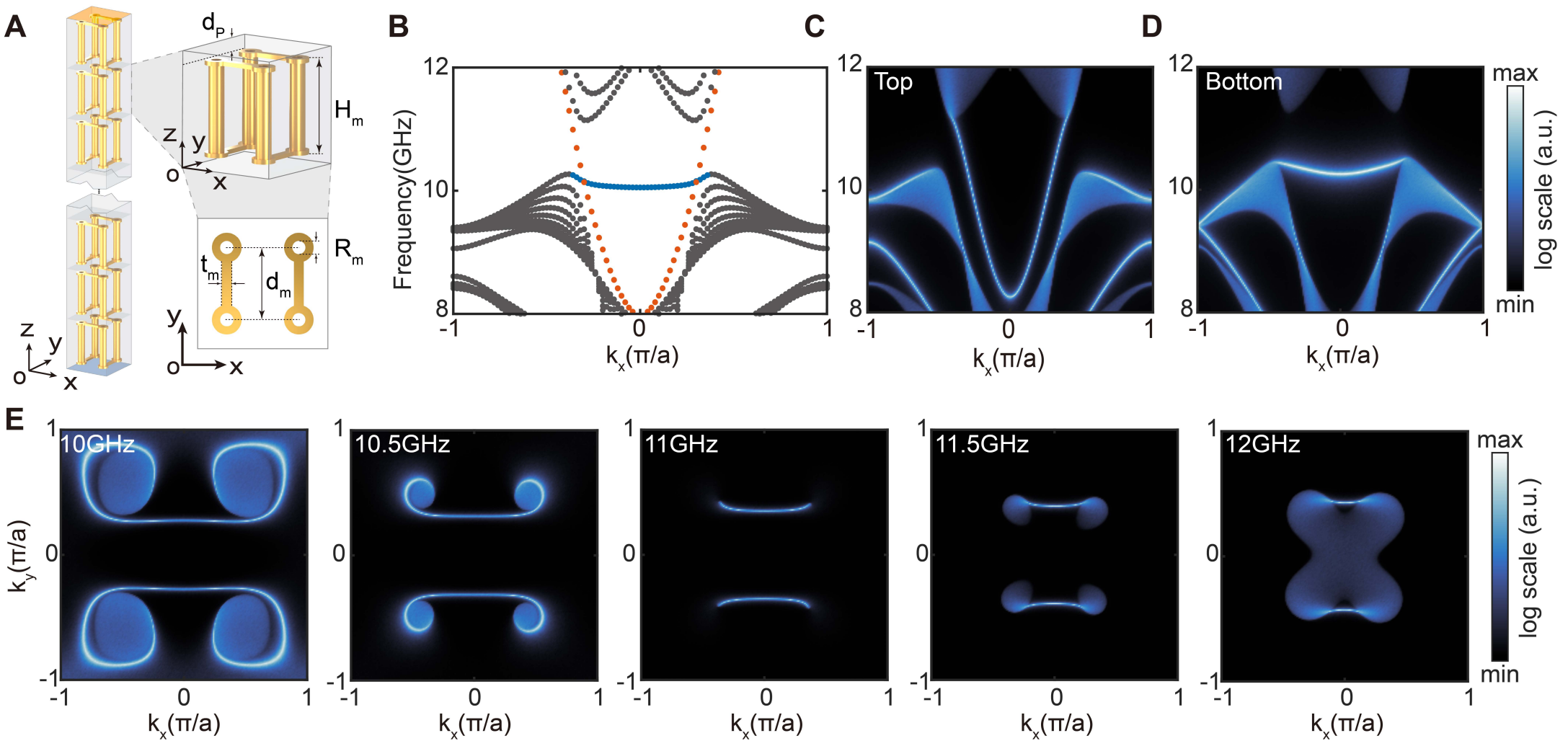

**Fig. 4. SDOS on a Semi-infinite 3D Weyl Metamaterial.** (a) Schematic of the saddle-shaped metamaterial, which behaves as an ideal Weyl system, with PEC boundary conditions at both the top and the bottom. (b) Band structure of a 9-cell supercell obtained by the SCM. (c, d) SDOS spectra calculated by LIM, with PEC boundaries at the top and bottom, respectively. (e) Evolution of the surface Fermi arcs in the SDOS equifrequency contours on the top surface from 10 to 12 GHz.

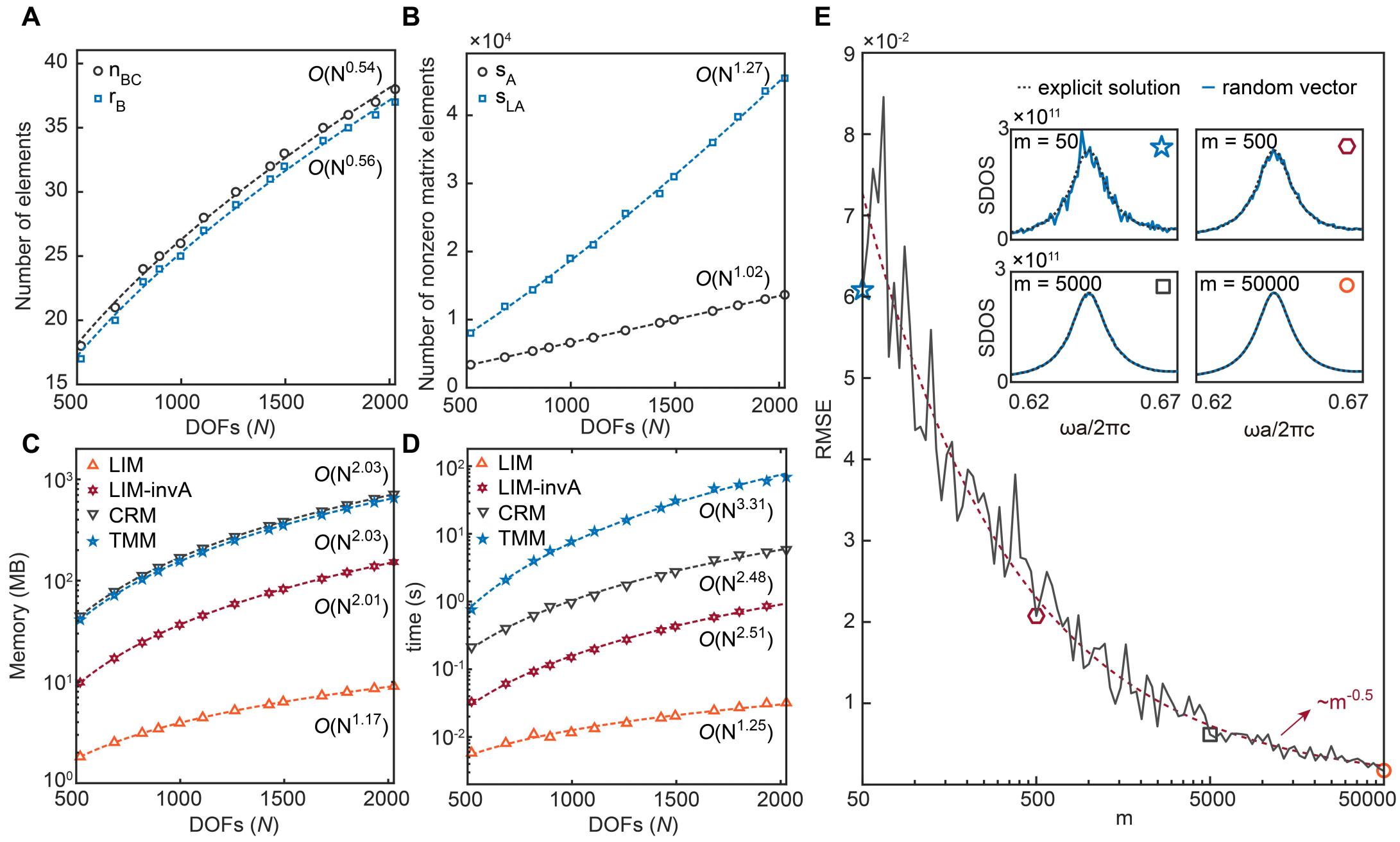

**Fig. 5. Computing Efficiency and Accuracy of the LIM Algorithm.** (a, b) Number of the $n_{BC}$ and $r_B$, and of nonzero matrix elements $\boldsymbol{s}_i$ of $\boldsymbol{A}$ and $\boldsymbol{L_A}$, in the 2D system, each fitted by a power function. (c, d) Average memory usage and computational time versus DOFs, obtained at 51 frequency points using LIM, LIM-invA, CRM, and TMM. For each algorithm, the growth rate with $N$ is fitted to a power law and marked to the right of the curve. (e) Error analysis of the stochastic trace estimator: RMSE versus the number of random vectors $m$, decreasing as $1/\sqrt{m}$. Insets show the SDOS versus frequency at $k_x = -0.8\pi/a$ for different $m$, illustrating the accuracy of the calculation.

**Table 1. Computational cost comparison of the four target functions in a 3D semi-infinite Weyl system with DOFs 4643.** Time and workspace memory ratios are normalized to the LIM baseline.

| Target Function | LIM | LIM-invA | CRM | TMM |
|---|---|---|---|---|
| **Avg Time (s)** | 2.5974 | 11.8334 | 38.1799 | 478.5181 |
| **Time Ratio** | 1 | 4.56 | 14.7 | 184.23 |
| **Workspace Memory (MB)** | 130.1375 | 582.0454 | 2214.2839 | 2032.0552 |
| **Workspace Memory Ratio** | 1 | 4.47 | 17.01 | 15.61 |

# Supplementary Materials:

## Low-Rank Iteration for Efficient Calculations of Topological Surface States

Kaiqi Liu[1], Changsheng He[3], Zihan Xie[1], Shulin Sun[1,3,4], Lei Zhou[2,3,4], Shaojie Ma[1,3,4*]

[1]College of Future Information Technology, Fudan University, Shanghai 200433, China

[2]Shanghai Key Laboratory of Metasurfaces for Light Manipulation, Shanghai 200433, China

[3]State Key Laboratory of Surface Physics and Department of Physics, Fudan University, Shanghai 200433, China

[4]Key Laboratory of Micro and Nano Photonic Structures (MOE), Fudan University, Shanghai 200433, China

[*]Correspondence: phzhou@fudan.edu.cn (L.Z.) and shaojiema@fudan.edu.cn (S.M.)

## This file includes:

1. Detailed Derivation of the LIM Algorithm
2. Discussion of the Metric $\boldsymbol{Q}$ in the SDOS calculation
3. Pseudocode of the LIM for representative partially infinite systems

# 1. Detailed Derivation of the LIM Algorithm

Starting from the block Riccati equation given in Eq. (6) of the main text, we have:

$$S = -A^{-1}C - A^{-1}B \cdot S^2 \tag{S1}$$

To proceed, we exploit the low-rank structure of the coupling matrices $\boldsymbol{B}$ and $\boldsymbol{C}$. Specifically, these matrices can be factorized as:

$$S = LV_C^\dagger,\ B = U_B V_B^\dagger,\ C = U_C V_C^\dagger \tag{S2}$$

Substituting Eq. (S2) into Eq. (S1) and rearranging the resulting terms yields:

$$LV_C^\dagger = -A^{-1}U_C V_C^\dagger - A^{-1}U_B V_B^\dagger (LV_C^\dagger)^2 \tag{S3}$$

Right-multiplying both sides of Eq. (S3) by $V_C$, we obtain:

$$L \cdot (V_C^\dagger V_C) = -A^{-1}U_C \cdot (V_C^\dagger V_C) - A^{-1} \cdot U_B V_B^\dagger \cdot LV_C^\dagger L \cdot (V_C^\dagger V_C) \tag{S4}$$

Here, the common factor $V_C^\dagger V_C$ is invertible and can therefore be eliminated completely, yielding:

$$L = -A^{-1}U_C - A^{-1}U_B \cdot V_B^\dagger L \cdot V_C^\dagger L \tag{S5}$$

Next, left-multiplying Eq. (S5) by $V_C^\dagger$ and extracting $V_C^\dagger L$, we arrive at:

$$[I + (V_C^\dagger A^{-1}U_B) \cdot V_B^\dagger L] \cdot V_C^\dagger L = -(V_C^\dagger A^{-1}U_C) \tag{S6}$$

From this, we can get:

$$V_C^\dagger L = -[I + (V_C^\dagger A^{-1}U_B) \cdot V_B^\dagger L]^{-1} \cdot (V_C^\dagger A^{-1}U_C) \tag{S7}$$

Reversing the steps from Eq. (S7) back to Eq. (S5) and rearranging the results, we arrive at:

$$L = -A^{-1}U_C + A^{-1}U_B \cdot V_B^\dagger L \cdot [I + (V_C^\dagger A^{-1}U_B) \cdot V_B^\dagger L]^{-1} \cdot (V_C^\dagger A^{-1}U_C) \tag{S8}$$

By invoking the following matrix identity:

$$(I + PQ)^{-1} \cdot P = P \cdot (I + QP)^{-1} \tag{S9}$$

With $P = V_B^\dagger L$ and $Q = V_C^\dagger A^{-1}U_B$, Eq. (S8) can be recast into the following form:

$$L = -A^{-1}U_C + A^{-1}U_B \cdot [I + (V_B^\dagger L) \cdot V_C^\dagger A^{-1}U_B]^{-1} \cdot V_B^\dagger L \cdot (V_C^\dagger A^{-1}U_C) \tag{S10}$$

Finally, by right-multiplying both sides of Eq. (S10) by $V_B^\dagger$, we obtain the self-consistent equation corresponding to Eq. (9) in the main text.

$$V_B{}^\dagger L = -V_B{}^\dagger A^{-1} U_C + V_B{}^\dagger A^{-1} U_B \cdot [I + (V_B{}^\dagger L) \cdot V_C^\dagger A^{-1} U_B]^{-1} \cdot V_B{}^\dagger L \cdot (V_C^\dagger A^{-1} U_C) \quad \text{(S11)}$$

In the computation, the following initialization matrices must be evaluated, along with the definition of $\tilde{\boldsymbol{L}}$:

$$\begin{gathered} \tilde{L} = V_B{}^\dagger L \\ A_{U_B} = A^{-1} U_B, A_{U_C} = A^{-1} U_C \\ V_{BAB} = V_B{}^\dagger A^{-1} U_B = V_B{}^\dagger A_{U_B},\ V_{BAC} = V_B{}^\dagger A^{-1} U_C = V_B{}^\dagger A_{U_C} \\ V_{CAB} = V_C{}^\dagger A^{-1} U = V_C{}^\dagger A_{U_B},\ V_{CAC} = V_B{}^\dagger A^{-1} U_C = V_B{}^\dagger A_{U_C} \end{gathered} \quad \text{(S12)}$$

Based on these definitions, the self-consistent equation for $\tilde{\boldsymbol{L}}$ can be written as:

$$\tilde{L} = -V_{BAC} + V_{BAB} \cdot \left[I + \tilde{L} \cdot V_{CAB}\right]^{-1} \cdot \tilde{L} \cdot V_{CAC} \quad \text{(S13)}$$

This self-consistent equation can be solved iteratively, with the iteration performed entirely in the low-rank subspace.

To obtain the explicit expression for $\boldsymbol{G}_0$ efficiently, we invoke the Sherman–Morrison–Woodbury (SMW) formula, which reads:

$$(A + UV^\dagger)^{-1} = A^{-1} - A^{-1} U \cdot (I + V^\dagger A^{-1} U)^{-1} \cdot V^\dagger A^{-1} \quad \text{(S14)}$$

The above form is valid provided that $A \in \mathbb{C}^{N \times N}$ and $I + V^\dagger A^{-1} U$ are invertible, and that $U, V \in \mathbb{C}^{N \times k}$. The formula can be verified directly by matrix multiplication. Substituting into Eq. (8) of the main text and using the low-rank decomposition of the coupling matrices, we obtain:

$$\begin{aligned} G_0 &= (A + BS)^{-1} \\ &= (A + U_B \cdot V_B{}^\dagger L V_C{}^\dagger)^{-1} \\ &= A^{-1} - A^{-1} U_B \cdot [I + V_B{}^\dagger L V_C{}^\dagger \cdot A^{-1} \cdot U_B]^{-1} \cdot V_B^\dagger L V_C{}^\dagger \cdot A^{-1} \end{aligned} \quad \text{(S15)}$$

Here, using the SMW formula, we define $U = U_B$ and $V = V_B^\dagger L V_C^\dagger$. Substituting the above definitions, we find that $\boldsymbol{G}_0$ depends entirely on $\tilde{\boldsymbol{L}}$:

$$G_0 = A^{-1} - A_{U_B} \cdot \left[I + \tilde{L} \cdot V_{CAB}\right]^{-1} \cdot \tilde{L} \cdot V_C{}^\dagger \cdot A^{-1} \quad \text{(S16)}$$

Likewise, the transfer matrix $\boldsymbol{S}$ can also be expressed in terms of $\tilde{\boldsymbol{L}}$ from Eq. (S10):

$$S = LV_C^{\dagger} = -A_{U_C} \cdot V_C^{\dagger} + A_{U_B} \cdot \left[I + \tilde{L} \cdot V_{CAB}\right]^{-1} \cdot \tilde{L} \cdot V_{CAC} \cdot V_C^{\dagger} \qquad \text{(S17)}$$

When the dispersion relation of the surface states is the only quantity of interest, the SDOS spectrum can be obtained by applying the stochastic trace estimator to Eq. (S16). If, however, the field distribution associated with the surface excitation must be computed rigorously, the inverse of $\boldsymbol{A}$ must be evaluated exactly according to Eq. (S16) to obtain $\boldsymbol{G}_0$, after which the transfer matrix $\boldsymbol{S}$ is obtained via Eq. (S17). The Green's function field distribution of the entire system is then given rigorously by $\boldsymbol{G}_n = \boldsymbol{S}^n \cdot \boldsymbol{G}_0$.

## 2. Discussion of the Metric $\boldsymbol{Q}$ in the SDOS calculation

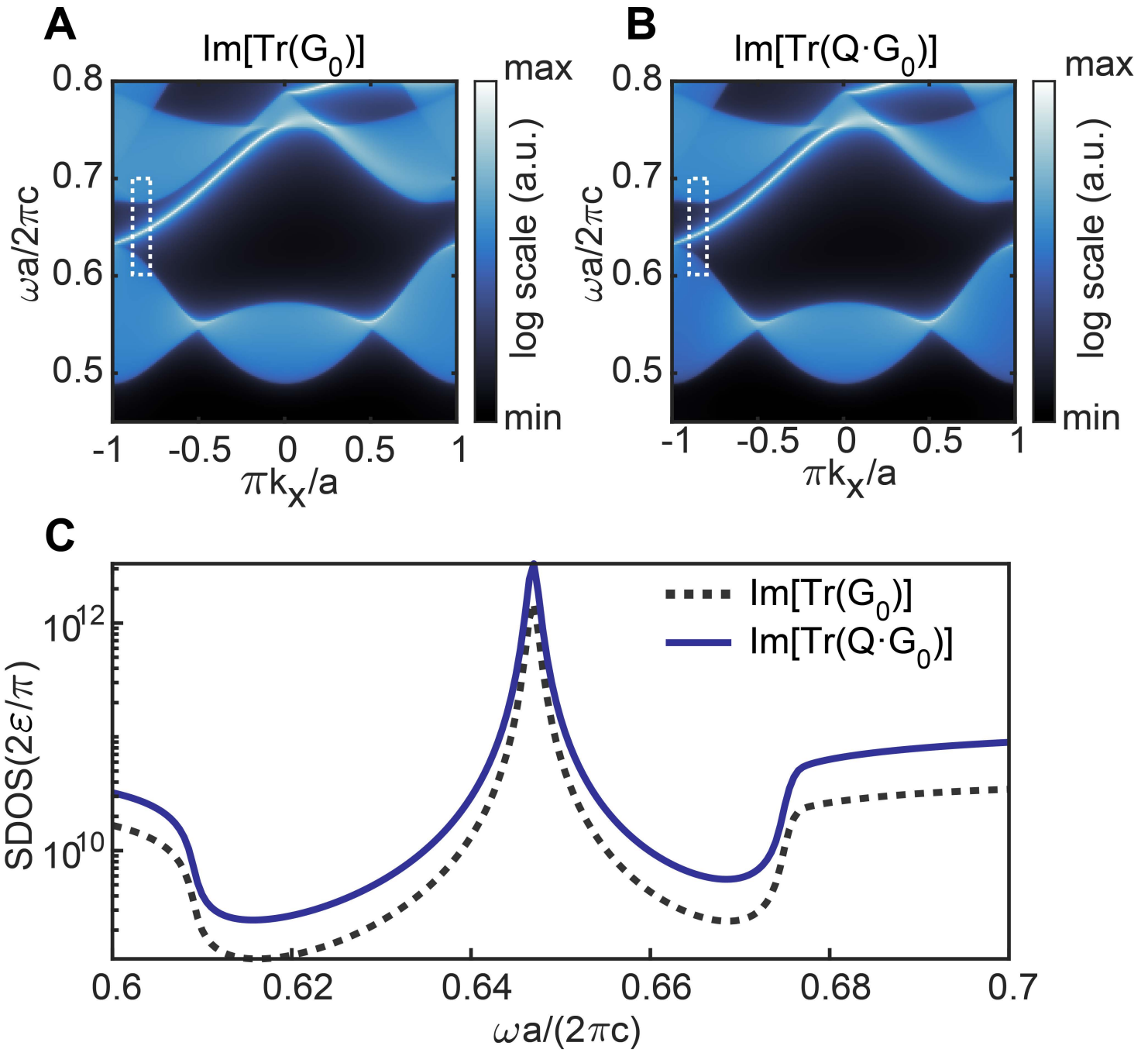

**Fig. S1. Comparison of the SDOS with and without the metric $\boldsymbol{Q}$.** (a) Simplified SDOS spectrum, with the formula $\mathrm{Im}(\mathrm{trace}(\boldsymbol{G}_0))$. (b) Rigorous SDOS spectrum, with the formula $\mathrm{Im}(\mathrm{trace}(\boldsymbol{Q}(\mathbf{r})\boldsymbol{G}_0))$. (c) Comparison of the SDOS with and without $\boldsymbol{Q}$ at $k_x = -0.8\pi/a$. Both spectra are calculated by LIM for the photonic crystal shown in Fig. 2 of the main text.

Strictly speaking, the SDOS result is given by Eq. (3), which depends not only on the surface Green's-function matrix $\boldsymbol{G}_0$ but also on the metric term $\boldsymbol{Q}(\mathbf{r})$. Once the metric is included, Eq. (11) of the main text can still be evaluated with the stochastic trace estimator, so the leading-order computational cost is unchanged. In practice, however, the metric itself can be rather complicated: in materials containing gyromagnetic terms, for example, $\boldsymbol{Q}(\mathbf{r})$ is a 3×3 tensor at every node, which complicates both its extraction and its application. In this work we therefore neglect $\boldsymbol{Q}(\mathbf{r})$ and directly use the imaginary part of the trace of $\boldsymbol{G}_0$ to locate the modes. The metric-weighted and unweighted spectra are not equal in a strict mathematical sense.

The SDOS peaks, however, arise from the divergent response of the system at the mode frequencies under a given excitation, and the same divergence occurs at the mode frequencies for both $\boldsymbol{G}_0$ and $\boldsymbol{Q} \cdot \boldsymbol{G}_0$. In electromagnetic systems, the two correspond respectively to the electric field and the electric displacement field, whose divergences likewise occur at the mode frequencies. Hence, provided that $\boldsymbol{Q}(\mathbf{r})$ is non-singular, the metric affects only the excitation strength and does not shift the mode positions.

To assess the influence of omitting $\boldsymbol{Q}(\mathbf{r})$, we calculated both the unweighted spectrum, $\mathrm{Im}(\mathrm{trace}(\boldsymbol{G}_0))$, and the weighted spectrum, $\mathrm{Im}(\mathrm{trace}(\boldsymbol{Q}(\mathbf{r})\boldsymbol{G}_0))$, for the 2D photonic crystal shown in **Fig. 2** of the main text. As shown in **Figs. S1(a-b)**, both approaches clearly identify the same surface state dispersion. Furthermore, **Fig. S1(c)** compares the two spectra over the frequency range $\omega = (0.6{\sim}0.7) \cdot a/2\pi c$ at $k_x = -0.8\pi/a$. Although the two approaches yield noticeably different absolute spectral intensities, their peak positions and overall line shapes are nearly identical, indicating the same surface mode positions. Therefore, omitting $\boldsymbol{Q}(\mathbf{r})$ does not affect the identification of the surface state position or its distinction from the bulk states in the present system.

This insensitivity can be understood from the pole structure of the Green's function: weighting the spectrum in post-processing changes the spectral weight of each mode without changing its resonance poles. Since the Green's function diverges at a mode in the limit $i \cdot \eta \rightarrow 0$, the peak position is fixed by its pole alone. Provided that no material parameter diverges at the same frequency, neither the spatial inhomogeneity nor the frequency dispersion of $\boldsymbol{Q}(\mathbf{r})$ can shift it. The two definitions therefore give the same peak positions and generally similar line shapes, while their absolute intensities may differ noticeably. An exception arises in extreme cases: at the resonance frequency of a lossless material, the material parameters, and hence the metric itself, can diverge, so that the dispersion requires additional care. For instance, the frequency-dependent terms can be linearized by introducing auxiliary variables, so that the response at that frequency can be evaluated correctly.

In summary, the analysis in this work focuses on the spectral positions and dispersions of the surface states, as well as their distinction from the bulk states, rather than on the absolute calibration of the SDOS. We also note that the peak intensity of the numerical spectrum depends on the artificially introduced imaginary component $i \cdot \eta$. Therefore, the spectral intensities in this work are primarily used for relative comparison and surface-state identification.

# 3. Pseudocode of the LIM for Representative Partially Infinite Systems

In this chapter, we present the pseudocode of the LIM for four types of partially infinite systems. These four cases cover most scenarios involving partially infinite systems, and the LIM for more complex systems can also be derived by reference to these cases.

## 3.1 LIM for a Bare Semi-Infinite Crystal

This section presents the LIM for a bare semi-infinite crystal, which corresponds to the system in **Fig. 2** and **Fig. 4** of the main text. The operator matrix $\boldsymbol{M}$ of this system can be expressed as:

$$\boldsymbol{M}=\left(\begin{array}{c|cccc} \boldsymbol{A} & \boldsymbol{B} & & & \\ \hline \boldsymbol{C} & \boldsymbol{A} & \boldsymbol{B} & & \\ & \boldsymbol{C} & \boldsymbol{A} & \boldsymbol{B} & \\ & & \ddots & \ddots & \ddots \end{array}\right) \tag{S18}$$

A detailed discussion of this system is given in the main text.

The following is the detailed pseudo-code of the algorithm 1:

| **Algorithm 1:** LIM for a bare semi-infinite crystal (Fig. 2 and Fig. 4 in the main text) |
|---|
| **Initialization:**<br>A, B, C<br>maximum iterations > 0, error tolerance > 0, number of random vectors > 0<br>**Output:**<br>trZ00 (trace of Z00)<br>1. **// Step 1: Low-rank truncation of *B* and *C* using QR decomposition**<br>2. $[Q_B, R_B] = \mathbf{qr}(B)$<br>3. $[Q_C, R_C] = \mathbf{qr}(C)$<br>4. $[U_B, V_B, U_C, V_C] = \mathbf{solveUV}(Q_B, R_B, Q_C, R_C)$<br>5. // Keep the columns $U_B$ and $U_C$ in $Q_B$ and $Q_C$ where the diagonal elements of $R_B$ and $R_C$ are non-zero, satisfying $B = U_B V_B^\dagger, C = U_C V_C^\dagger$<br>6. **// Step 2: Precompute quantities independent of the iteration** |

$AinvUB = A^{-1}U_B$
$AinvUC = A^{-1}U_C$
$P = V_B^{\dagger} \times AinvUC$
$Q = V_B^{\dagger} \times AinvUB$
$R = V_C^{\dagger} \times AinvUB$
$T = V_C^{\dagger} \times AinvUC$
**// Step 3: Solve for the $\tilde{L} = V_B' L$ using low-rank Riccati iteration**
$Lt = 0$
**for** $i$ = 1 **to** maximum iterations **do**:
    $Ltnew = -P + Q \times (I_{r_B} + Lt \times R)^{-1} \times (Lt \times T)$
    **if** $||Ltnew - Lt||_F / (||Lt||_F + eps) < \ error\ tolerance$ **then**
        **break**
    **end**
    $Lt = Ltnew$
**end**
**// Step 4: Precompute fixed matrices for Woodbury formula**
$VBLfinal = Lt$
$K = I_{r_B} + VBLfinal \cdot R$
$[LK,\ UK] \ = \ \mathbf{lu}(K)$
**// Step 5: Estimate $tr((A \ + \ B\,S)^{-1})$ using Hutchinson's method**
$z = \mathbf{randomVector}\,(m)$
// Generate $m$ Rademacher random vectors
$w = A^{-1}z$
$s = UK^{-1}(LK^{-1} \times VBLfinal \times V_C^{\dagger} w)$
$Z00z \ = \ w - AinvUB \times s$
$sumtr = \mathbf{sum}\,(z^{\dagger} \times Z00z)$
//Calculate each column's $z_i^{\dagger} \times Z00 \times z_i$
$trZ00 = sumtr/m$
**return** trZ00

Noted that the inverse matrix operator $\boldsymbol{A^{-1}}$ here is used only to express the mathematical formula clearly. In the calculation, LU decomposition is employed to obtain the inverse matrix and its products $\boldsymbol{A^{-1}U_B}$, $\boldsymbol{A^{-1}U_C}$ and $\boldsymbol{A^{-1}z}$.

## 3.2 LIM for a Semi-Infinite Crystal with a Modified Boundary Block

This section presents the LIM for a semi-infinite crystal with a modified boundary block, which corresponds to the system in Fig. 3(a) of the main text involving additional cladding layers. The operator matrix $\boldsymbol{M}$ of this system can be expressed as:

$$
\boldsymbol{M}=\left(\begin{array}{c|cccc}
\boldsymbol{A}_0 & \boldsymbol{B}_0 & & & \\
\hline
\boldsymbol{C}_0 & \boldsymbol{A}_1 & \boldsymbol{B}_1 & & \\
 & \boldsymbol{C}_1 & \boldsymbol{A}_1 & \boldsymbol{B}_1 & \\
 & & \ddots & \ddots & \ddots
\end{array}\right) \tag{S19}
$$

The difference from algorithm 1 lies in the additional treatment required for the coupling of surface Green's functions.

In this modified semi-infinity system, obtaining the final response requires considering two distinct sets of Green's functions simultaneously: the bulk surface Green's function $\boldsymbol{z}_{11}$ and the boundary surface Green's function $\boldsymbol{z}_{00}$. They are defined respectively as:

$$
\boldsymbol{z}_{11}=(\boldsymbol{A}_1+\boldsymbol{B}_1\boldsymbol{S})^{-1},\ \boldsymbol{z}_{00}=(\boldsymbol{A}_0-\boldsymbol{B}_0\boldsymbol{z}_{11}\boldsymbol{C}_0)^{-1} \tag{S20}
$$

Both Green's functions can be converted into explicit expressions via Woodbury transformations. The core expressions of the two Woodbury transformations are as follows:

$$
\boldsymbol{z}_{11}=\boldsymbol{A}_1^{-1}-\boldsymbol{A}_1^{-1}\boldsymbol{U}_{\boldsymbol{B}_1}\left[\boldsymbol{I}+\tilde{\boldsymbol{L}}\boldsymbol{V}_{\boldsymbol{C}_1}^{\dagger}\boldsymbol{A}_1^{-1}\boldsymbol{U}_{\boldsymbol{B}_1}\right]^{-1}\tilde{\boldsymbol{L}}\boldsymbol{V}_{\boldsymbol{C}_1}^{\dagger}\boldsymbol{A}_1^{-1} \tag{S21}
$$

and

$$
\boldsymbol{z}_{00}=\boldsymbol{A}_0^{-1}+\boldsymbol{A}_0^{-1}\boldsymbol{U}_{\boldsymbol{B}_0}\left[\boldsymbol{I}-\tilde{\boldsymbol{L}}_{00}\boldsymbol{V}_{\boldsymbol{C}_0}^{\dagger}\boldsymbol{A}_0^{-1}\boldsymbol{U}_{\boldsymbol{B}_0}\right]^{-1}\tilde{\boldsymbol{L}}_{00}\boldsymbol{V}_{\boldsymbol{C}_0}^{\dagger}\boldsymbol{A}_0^{-1} \tag{S22}
$$

where $\tilde{\boldsymbol{L}}_{00}=\boldsymbol{V}_{\boldsymbol{B}_0}^{\dagger}\boldsymbol{z}_{11}\boldsymbol{U}_{\boldsymbol{C}_0}$. The final SDOS can be expressed as a trace involving $\boldsymbol{Z}_{00}$, which can likewise be obtained using the stochastic trace estimator:

$$
\operatorname{tr}(\boldsymbol{Z}_{00})=\frac{1}{m}\sum_{i=1}^{m}\boldsymbol{z}_i^{\dagger}\boldsymbol{Z}_{00}\boldsymbol{z}_i \tag{S23}
$$

The following is the detailed pseudo-code of the algorithm 2:

**Algorithm 2:** LIM for a semi-infinite crystal with a boundary block (Fig. 3a in the main text)

**Initialization:**

A0, B0, C0, A1, B1, C1

maximum iterations > 0, error tolerance > 0, number of random vectors > 0

**Output:**

trZ00 (trace of Z00)

// **Step 1: Low-rank truncation of $\boldsymbol{B}_*$ and $\boldsymbol{C}_*$ using QR decomposition**

$[Q_{B_0},R_{B_0}]=\mathbf{qr}(B_0)$

$[Q_{C_0},R_{C_0}]=\mathbf{qr}(C_0)$

$[Q_{B_1}, R_{B_1}] = \textbf{qr}(B_1)$

$[Q_{C_1}, R_{C_1}] = \textbf{qr}(C_1)$

$[U_{B_0}, V_{B_0}, U_{C_0}, V_{C_0}, U_{B_1}, V_{B_1}, U_{C_1}, V_{C_1}] = \textbf{solveUV}(Q_{B_0}, R_{B_0}, Q_{C_0}, R_{C_0}, Q_{B_1}, R_{B_1}, Q_{C_1}, R_{C_1})$

// Keep the columns $U*$ in $Q*$ where the diagonal elements of $R*$ are non-zero, satisfying $B_* = U_{B_*} V_{B_*}^{\dagger}, C_* = U_{C_*} V_{C_*}^{\dagger}$

**// Step 2: Precompute quantities independent of the iteration**

$A1invUB1 = A_1^{-1} U_{B_1}$

$A1invUC1 = A_1^{-1} U_{C_1}$

$A1invUC0 = A_1^{-1} U_{C_0}$

$A0invUB0 = A_0^{-1} \cdot U_{B_0}$

$P = V_{B_1}^{\dagger} \times A1invUC1$

$Q = V_{B_1}^{\dagger} \times A1invUB1$

$R = V_{C_1}^{\dagger} \times A1invUB1$

$T = V_{C_1}^{\dagger} \times A1invUC1$

**// Step 3: Solve for the $\tilde{L}_1 = V_{B_1}' L$ using low-rank Riccati iteration**

$Lt = 0$

**for** $i$ = 1 **to** maximum iterations **do**:

  $Ltnew = -P + Q \cdot [(I_{B_1} + Lt \times R)^{-1} \times (Lt \times T)]$

  **if** $||Ltnew - Lt||_F / (||Lt||_F + eps) < error\ tolerance$ **then**

    **break**

  **end**

  $Lt = Ltnew$

**end**

**// Step 4: Precompute fixed matrices for Woodbury formula**

$VBLfinal = Lt$

$Kbulk = I_{r_{B_1}} + Lt \times R$

$Sbulk = Kbulk^{-1} \times [Lt \times (V_{C_1}' \times A1invUC0)]$

$z11UC0 = A1invUC0 - A1invUB1 \times Sbulk$

**// Step 5: Construct the boundary Green's function using the Woodbury formula**

$Lt00 = V_{B_0}^{t} \times z11UC0$

$Kbd = I - Lt00 \times V_{C_0}^{t} \times A0invUB0$

**// Step 6: Estimate $\mathrm{tr}(Z_{00})$ using Hutchinson's method**

$z = \textbf{randomVector}(m)$

// Generate $m$ Rademacher random vectors

$w = A_0^{-1} z$

$s = Kbd^{-1} \times Lt00 \times (V_{C0}^{t} \times w)$

$Z00z = w + A0invUB0 \times s$

$sumtr = \textbf{sum}(z' \times Z00z)$

//Calculate each column's $z_i' \times (Z00 \times z_i)$

$trZ00 = sumtr/m$

**return** trZ00

## 3.3 LIM for Two Semi-Infinite Crystals Interfaced with Each Other

This section presents the LIM for two semi-infinite crystals separated by another crystal slab, which corresponds to the photonic crystal in Fig. 3(d) and the phononic crystal in Fig. 3(j) of the main text. The operator matrix $\boldsymbol{M}$ of both systems can be expressed as:

$$\boldsymbol{M}=\left(\begin{array}{cccc:ccc} \ddots & \ddots & \ddots & & & & \\ & \boldsymbol{C}_{11} & \boldsymbol{A}_{11} & \boldsymbol{B}_{11} & & & \\ & & \boldsymbol{C}_{11} & \boldsymbol{A}_{11} & \boldsymbol{B}_{12} & & \\ \hdashline & & & \boldsymbol{C}_{21} & \boldsymbol{A}_{22} & \boldsymbol{B}_{22} & \\ & & & & \boldsymbol{C}_{22} & \boldsymbol{A}_{22} & \boldsymbol{B}_{22} \\ & & & & & \ddots & \ddots & \ddots \end{array}\right) \quad \text{(S24)}$$

For these systems, the interface concatenation no longer includes an independent surface layer. Instead, it is composed of the last $\boldsymbol{A}_{11}$ layer on the left and the first $\boldsymbol{A}_{22}$ layer on the right.

The Green's functions on both sides $\boldsymbol{z}_L$ and $\boldsymbol{z}_R$ respectively satisfy:

$$\begin{cases} \boldsymbol{z}_L=(\boldsymbol{A}_{11}+\boldsymbol{C}_{11}\boldsymbol{S}_L)^{-1} \\ \boldsymbol{z}_R=(\boldsymbol{A}_{22}+\boldsymbol{B}_{22}\boldsymbol{S}_R)^{-1} \end{cases} \quad \text{(S25)},$$

where $\boldsymbol{S}_L$ and $\boldsymbol{S}_R$ represent the transfer matrices. Therefore, the intermediate effective matrix should be defined as the block Green's function on the two layers of the interface:

$$\boldsymbol{M}_{eff}=\begin{bmatrix} \boldsymbol{Z}_L^{-1} & \boldsymbol{B}_{12} \\ \boldsymbol{C}_{21} & \boldsymbol{Z}_R^{-1} \end{bmatrix} \quad \text{(S26)}$$

And the Green's function in the middle of the interface is strictly defined as:

$$\boldsymbol{Z}_{eff}=\boldsymbol{M}_{eff}^{-1}=\begin{bmatrix} \boldsymbol{G}_{11} & \boldsymbol{G}_{12} \\ \boldsymbol{G}_{21} & \boldsymbol{G}_{22} \end{bmatrix} \quad \text{(S27)}$$

This expression can also be converted into an explicit form via Woodbury transformations. Specially, the Woodbury factor in the pseudo-code satisfies:

$$\boldsymbol{U}_{int}=\begin{pmatrix}\boldsymbol{U}_{\boldsymbol{C}_{11}} & 0 & \boldsymbol{U}_{\boldsymbol{B}_{12}} & 0\\ 0 & \boldsymbol{U}_{\boldsymbol{B}_{22}} & 0 & \boldsymbol{U}_{\boldsymbol{C}_{21}}\end{pmatrix},\ \boldsymbol{V}_{int}=\begin{pmatrix}\boldsymbol{V}_{\boldsymbol{B}_{11}} & 0 & 0 & \boldsymbol{V}_{\boldsymbol{C}_{21}}\\ 0 & \boldsymbol{V}_{\boldsymbol{C}_{22}} & \boldsymbol{V}_{\boldsymbol{B}_{12}} & 0\end{pmatrix},$$
$$\widetilde{\boldsymbol{L}}_{int}=-\,\mathbf{blkdiag}\left(\widetilde{\boldsymbol{L}}_{11},\widetilde{\boldsymbol{L}}_{22},\boldsymbol{I}_{\boldsymbol{B}_{12}},\boldsymbol{I}_{\boldsymbol{C}_{21}}\right) \tag{S28}$$

and the surface of Woodbury transformations is:

$$\boldsymbol{Z}_{int}=\boldsymbol{A}_{int}^{-1}+\boldsymbol{A}_{int}^{-1}\boldsymbol{U}_{int}\cdot\left[\boldsymbol{I}-\widetilde{\boldsymbol{L}}_{int}\boldsymbol{V}_{int}^{\dagger}\boldsymbol{A}_{int}^{-1}\boldsymbol{U}_{int}\right]^{-1}\cdot\widetilde{\boldsymbol{L}}_{int}\boldsymbol{V}_{int}^{\dagger}\cdot\boldsymbol{A}_{int}^{-1} \tag{S29}$$

where $\boldsymbol{A}_{int}=\mathbf{blkdiag}\left(\boldsymbol{A}_{11},\boldsymbol{A}_{22}\right)$.

The final SDOS can be expressed as a trace involving $\boldsymbol{Z_{int}}$, which can likewise be obtained using the stochastic trace estimation:

$$\mathrm{tr}\left(\boldsymbol{Z}_{int}\right)=\frac{1}{m}\sum_{i=1}^{m}\boldsymbol{z}_i^{\dagger}\boldsymbol{Z}_{int}\boldsymbol{z}_i \tag{S30}$$

The following is the detailed pseudo-code of the algorithm 3:

**Algorithm 3:** LIM for two semi-infinite crystals interfaced with each other (Fig.3d in the main text)

**Initialization:**

A11, B11, C11, A22, B22, C22, B12, C21

maximum iterations > 0, error tolerance > 0, number of random vectors > 0

**Output:**

$\mathbf{tr}\left(\boldsymbol{Z}_{\mathrm{mid}}\right)$ (trace of $\boldsymbol{Z}_{\mathrm{mid}}$)

**// Step 1: Low-rank truncation of $B_*$ and $C_*$ using QR decomposition**

$[Q_{B_{11}},R_{B_{11}}]=\mathbf{qr}\left(B_{11}\right)$

$[Q_{C_{11}},R_{C_{11}}]=\mathbf{qr}\left(C_{11}\right)$

$[Q_{B_{22}},R_{B_{22}}]=\mathbf{qr}\left(B_{22}\right)$

$[Q_{C_{22}},R_{C_{22}}]=\mathbf{qr}\left(C_{22}\right)$

$[Q_{B_{12}},R_{B_{12}}]=\mathbf{qr}\left(B_{12}\right)$

$[Q_{C_{21}},R_{C_{21}}]=\mathbf{qr}\left(C_{21}\right)$

$[U_{B_{11}},V_{B_{11}},U_{B_{22}},V_{B_{22}},U_{B_{12}},V_{B_{12}}]=\mathbf{solveUV}\left(Q_{B_{11}},R_{B_{11}},Q_{B_{22}},R_{B_{22}},Q_{B_{12}},R_{B_{12}}\right)$

$[U_{C_{11}},V_{C_{11}},U_{C_{22}},V_{C_{22}},U_{C_{21}},V_{C_{21}}]=\mathbf{solveUV}\left(Q_{C_{11}},R_{C_{11}},Q_{C_{22}},R_{C_{22}},Q_{C_{21}},R_{C_{21}}\right)$

// Keep the columns $U_*$ in $Q_*$ where the diagonal elements of $R_*$ are non-zero, satisfying $B_*=U_{B_*}V_{B_*}{}^{\dagger},C_*=U_{C_*}V_{C_*}{}^{\dagger}$

**// Step 2: Precompute quantities independent of the iteration**

$A11invUB11=A_{11}^{-1}U_{B_{11}}$

$A11invUC11=A_{11}^{-1}U_{C_{11}}$

$A22invUB22 = A_{22}^{-1} U_{B_{22}}$
$A22invUC22 = A_{22}^{-1} U_{C_{22}}$
$Pleft = V_{C_{11}}^{\dagger} \times A11invUB11$
$Qleft = V_{C_{11}}^{\dagger} \times A11invUC11$
$Rleft = V_{B_{11}}^{\dagger} \times A11invUC11$
$Tleft = V_{B_{11}}^{\dagger} \times A11invUB11$
$Pright = V_{B_{22}}^{\dagger} \times A22invUC22$
$Qright = V_{B_{22}}^{\dagger} \times A22invUB22$
$Rright = V_{C_{22}}^{\dagger} \times A22invUB22$
$Tright = V_{C_{22}}^{\dagger} \times A22invUC22$
**// Step 3: Solve for the *L* using low-rank Riccati iteration**
$Ltleft = 0$
$Ltright = 0$
**for** $i$ = 1 **to** maximum iterations **do**:
    $Kleft = I + Ltleft \times Rleft$
    $LtLeftnew = -Pleft + Qleft \times [Kleft^{-1} \times (Ltleft \times Tleft)]$
    **if** $||LtLeftnew - LtLeft||_F / (||LtLeft||_F + eps) < error\ tolerance$ **then**
        **break**
    **end**
    $LtLeft = LtLeftnew$
**end**
**for** $i$ = 1 **to** maximum iterations **do**:
    $Kright = I + Ltright \times Rright$
    $Ltrightnew = -PLtRight + Qright \times [Kright^{-1} \times (Ltright \times Tright)]$
    **if** $||Ltrightnew - Ltright||_F / (||Ltright||_F + eps) < error\ tolerance$ **then**
        **break**
    **end**
    $Ltright = Ltrightnew$
**end**
**// Step 4: Precompute fixed matrices for Woodbury formula**
$Ubd = [U_{C_{11}}, 0, U_{B_{12}}, 0; 0, U_{B_{22}}, 0, U_{C_{21}}]$
$Vbd = [V_{B_{11}}, 0, 0, V_{C_{21}}; 0, V_{C_{22}}, V_{B_{12}}, 0]$
$Ltint = -$ **bikdiag** $(Ltleft, Ltright, I, I)$
$Aint =$ **blkdiag** $(A_{11}, A_{22})^{-1} \times U_{bd}$
$Kbd = I - Ltint \times (Vbd' \times Aint)$
**// Step 6: Estimate $\mathbf{tr}(\boldsymbol{Z}_{00})$ using Hutchinson's method**
$z =$ **randomVector** $(m)$
// Generate $m$ Rademacher random vectors
$w = [A_{11}^{-1} z(left),\ A_{22}^{-1} z(right)]$
$s = Kbd^{-1} \times Ltint \times (Vbd' \times w)$
$Zmidz = w + Aint \times s$
$sumtr =$ **sum** $(z' \times Z00z)$
//Calculate each column's $z_i^{\dagger} \times Z00 \times z_i$
$trZmid = sumtr / m$

```
return trZmid
```

## 3.4 LIM for Two Semi-Infinite Crystals Separated by Another Crystal Slab

This section presents the LIM for two semi-infinite crystals separated by another crystal slab, which corresponds to the system in Fig. 3(g) of the main text. The operator matrix M of this system can be expressed as:

$$\boldsymbol{M}=\left(\begin{array}{cccc:c:ccc}
\ddots & \ddots & \ddots & & & & & \\
 & \boldsymbol{C}_{11} & \boldsymbol{A}_{11} & \boldsymbol{B}_{11} & & & & \\
 & & \boldsymbol{C}_{11} & \boldsymbol{A}_{11} & \boldsymbol{B}_{01} & & & \\
\hdashline
 & & & \boldsymbol{C}_{10} & \boldsymbol{A}_{00} & \boldsymbol{B}_{02} & & \\
\hdashline
 & & & & \boldsymbol{C}_{20} & \boldsymbol{A}_{22} & \boldsymbol{B}_{22} & \\
 & & & & & \boldsymbol{C}_{22} & \boldsymbol{A}_{22} & \boldsymbol{B}_{22} \\
 & & & & & & \ddots & \ddots \quad \ddots
\end{array}\right) \tag{S31}$$

In this system, the bulk Green's function in both sides can be defined as:

$$\begin{cases} \boldsymbol{Z}_{11}=(\boldsymbol{A}_{11}-\boldsymbol{C}_{11}\boldsymbol{S}_{11})^{-1} \\ \boldsymbol{Z}_{22}=(\boldsymbol{A}_{22}-\boldsymbol{B}_{22}\boldsymbol{S}_{22})^{-1} \end{cases} \quad \text{(S32)},$$

and the surface Green's function satisfies:

$$\boldsymbol{Z}_{00}=(\boldsymbol{A}_{00}-\boldsymbol{C}_{10}\boldsymbol{Z}_{11}\boldsymbol{B}_{01}-\boldsymbol{B}_{02}\boldsymbol{Z}_{22}\boldsymbol{C}_{20})^{-1} \tag{S33}$$

This expression can also be converted into an explicit form via Woodbury transformations. The core expressions are as follows:

$$\begin{cases} \boldsymbol{z}_{11}=\boldsymbol{A}_{11}^{-1}-\boldsymbol{A}_{11}^{-1}\boldsymbol{U}_{\boldsymbol{B}_{11}}\cdot\left[\boldsymbol{I}+\tilde{\boldsymbol{L}}_{11}\boldsymbol{V}_{\boldsymbol{C}_{11}}^{\dagger}\boldsymbol{A}_{11}^{-1}\boldsymbol{U}_{\boldsymbol{B}_{11}}\right]^{-1}\cdot\tilde{\boldsymbol{L}}_{11}\boldsymbol{V}_{\boldsymbol{C}_{11}}^{\dagger}\cdot\boldsymbol{A}_{11}^{-1} \\ \boldsymbol{z}_{22}=\boldsymbol{A}_{22}^{-1}-\boldsymbol{A}_{22}^{-1}\boldsymbol{U}_{\boldsymbol{B}_{22}}\cdot\left[\boldsymbol{I}+\tilde{\boldsymbol{L}}_{22}\boldsymbol{V}_{\boldsymbol{C}_{22}}^{\dagger}\boldsymbol{A}_{22}^{-1}\boldsymbol{U}_{\boldsymbol{B}_{22}}\right]^{-1}\cdot\tilde{\boldsymbol{L}}_{22}\boldsymbol{V}_{\boldsymbol{C}_{22}}^{\dagger}\cdot\boldsymbol{A}_{22}^{-1} \end{cases} \quad \text{(S34)},$$

and the final surface Green's function is:

$$\boldsymbol{Z}_{00} = \boldsymbol{A}_{00}^{-1} + \boldsymbol{A}_{00}^{-1}[\boldsymbol{U}_{\boldsymbol{C}_{10}}\ \ \boldsymbol{U}_{\boldsymbol{B}_{02}}] \cdot \left[\boldsymbol{I} - \tilde{\boldsymbol{L}}_{00}\begin{bmatrix}\boldsymbol{V}_{\boldsymbol{C}_{10}}^{\dagger}\\ \boldsymbol{V}_{\boldsymbol{C}_{20}}^{\dagger}\end{bmatrix}\boldsymbol{A}_{00}^{-1}[\boldsymbol{U}_{\boldsymbol{B}_{01}}\ \ \boldsymbol{U}_{\boldsymbol{B}_{02}}]\right]^{-1} \cdot \tilde{\boldsymbol{L}}_{00}\begin{bmatrix}\boldsymbol{V}_{\boldsymbol{C}_{10}}^{\dagger}\\ \boldsymbol{V}_{\boldsymbol{C}_{20}}^{\dagger}\end{bmatrix} \cdot \boldsymbol{A}_{00}^{-1} \quad \text{(S35)},$$

where $\tilde{\boldsymbol{L}}_{00} = \mathbf{blkdiag}(\boldsymbol{V}_{\boldsymbol{B}_{01}}' \boldsymbol{Z}_{11}\boldsymbol{U}_{\boldsymbol{C}_{10}}\ \ \boldsymbol{V}_{\boldsymbol{B}_{02}}' \boldsymbol{Z}_{22}\boldsymbol{U}_{\boldsymbol{C}_{20}})$.

The final SDOS can be expressed as a trace involving $\boldsymbol{Z}_{00}$, which can likewise be obtained using the stochastic trace estimation:

$$\mathrm{tr}(\boldsymbol{Z}_{00}) = \frac{1}{m}\sum_{i=1}^{m} \boldsymbol{z}_i^{\dagger}\boldsymbol{Z}_{00}\boldsymbol{z}_i \quad \text{(S36)}$$

The following is the detailed pseudo-code of the algorithm 4:

**Algorithm 4:** LIM for two semi-infinite crystals separated by another crystal slab (Fig.3g in the main text)

**Initialization:**

A00, B01, C10, B02, C20, A11, B11, C11, A22, B22, C22

maximum iterations > 0, error tolerance > 0, number of random vectors > 0

**Output:**

trZ00 (trace of Z00)

**// Step 1: Low-rank truncation of $B_*$ and $C_*$ using QR decomposition**

$[Q_{B_{01}}, R_{B_{01}}] = \mathbf{qr}(B_{01})$

$[Q_{C_{10}}, R_{C_{10}}] = \mathbf{qr}(C_{10})$

$[Q_{B_{02}}, R_{B_{02}}] = \mathbf{qr}(B_{02})$

$[Q_{C_{20}}, R_{C_{20}}] = \mathbf{qr}(C_{20})$

$[Q_{B_{11}}, R_{B_{11}}] = \mathbf{qr}(B_{11})$

$[Q_{C_{11}}, R_{C_{11}}] = \mathbf{qr}(C_{11})$

$[Q_{B_{22}}, R_{B_{22}}] = \mathbf{qr}(B_{22})$

$[Q_{C_{22}}, R_{C_{22}}] = \mathbf{qr}(C_{22})$

$[U_{B_{01}}, V_{B_{01}}, U_{C_{10}}, V_{C_{10}}, U_{B_{02}}, V_{B_{02}}, U_{C_{20}}, V_{C_{20}}] = \mathbf{solveUV}(Q_{B_{01}}, R_{B_{01}}, Q_{C_{10}}, R_{C_{10}}, Q_{B_{02}}, R_{B_{02}}, Q_{C_{20}}, R_{C_{20}})$

$[U_{B_{11}}, V_{B_{11}}, U_{C_{11}}, V_{C_{11}}, U_{B_{22}}, V_{B_{22}}, U_{C_{22}}, V_{C_{22}}] = \mathbf{solveUV}(Q_{B_{11}}, R_{B_{11}}, Q_{C_{11}}, R_{C_{11}}, Q_{B_{22}}, R_{B_{22}}, Q_{C_{22}}, R_{C_{22}})$

// Keep the columns $U_*$ in $Q_*$ where the diagonal elements of $R_*$ are non-zero, satisfying

$$\begin{cases} B_* = U_{B_*}V_{B_*}^{\dagger} \\ C_* = U_{C_*}V_{C_*}^{\dagger} \end{cases}$$

**// Step 2: Precompute quantities independent of the iteration**

$A11invUB11 = A_{11}^{-1}U_{B_{11}}$

$A11invUC11 = A_{11}^{-1}U_{C_{11}}$

$A11invUC10 = A_{11}^{-1}U_{C_{10}}$

$A22invUB22 = A_{22}^{-1}U_{B_{22}}$

$A22invUC22 = A_{22}^{-1}U_{C_{22}}$

$A22invUC20 = A_{22}^{-1}U_{C_{20}}$

$Pleft = V_{B_{11}}' \times A11invUC11$

$Qleft = V'_{B_{11}} \times A11invUB11$

$Rleft = V'_{C_{11}} \times A11invUB11$

$Tleft = V'_{C_{11}} \times A11invUC11$

$Pright = V'_{B_{22}} \times A22invUC22$

$Qright = V'_{B_{22}} \times A22invUB22$

$Rright = V'_{C_{22}} \times A22invUB22$

$Tright = V'_{C_{22}} \times A22invUC22$

**// Step 3: Solve for the $V'_{B_*}L$ using low-rank Riccati iteration**

$Ltleft = 0$

$Ltright = 0$

**for** $i$ = 1 **to** maximum iterations **do**:

  $Kleft = I + Ltleft \times Rleft$

  $LtleftNew = -Pleft + Qleft \times [Kleft^{-1} \times (Ltleft \times Tleft)]$

  **if** $||LtleftNew - Ltleft||_F / (||Ltleft||_F + eps) <$ *error tolerance* **then**

    **break**

  **end**

  $Ltleft = LtleftNew$

**end**

**for** $i$ = 1 **to** maximum iterations **do**:

  $Kright = I + Ltright \times Rright$

  $LtrightNew = -Pright + Qright \times [Kright^{-1} \times (Ltright \times Tright)]$

  **if** $||LtrightNew - Ltright||_F / (||Ltright||_F + eps) <$ *error tolerance* **then**

    **break**

  **end**

  $Ltright = LtrightNew$

**end**

**// Step 4: Precompute fixed matrices for Woodbury formula**

$Kleft = I + Ltleft \times Rleft$

$Kright = I + Ltright \times Rright$

$ZleftUC10 = A11invUC10 - A11invUB11 \times \{Kleft^{-1} \times [Ltleft \times (V'_{C_{11}} \times A11invUC10)]\}$

$ZrightUC20 = A22invUC20 - A22invUB22 \times \{Kright^{-1} \times [Ltright \times (V'_{C_{22}} \times A22invUC20)]\}$

**// Step 5: Construct the boundary Green's function using the Woodbury formula**

$Ubd = [U_{B_{01}}, U_{B_{02}}]$

$Vbd = [V_{C_{10}}, V_{C_{20}}]$

$A00invUbd = A_{00}^{-1} \cdot Ubd$

$Hleft = V^{\dagger}_{B_{01}} \times ZleftUC10$

$Hright = V^{\dagger}_{B_{02}} \times ZrightUC20$

$Ltint = \textbf{blkdiag}(Hleft, Hright)$

$Kbd = I\text{-}Ltint \times (Vbd^{\dagger} \times A00invUbd)$

**// Step 6: Estimate $\mathrm{tr}(Z_{00})$ using Hutchinson's method**

$z = \textbf{randomVector}(m)$

// Generate $m$ Rademacher random vectors

$w = A_{00}^{-1} z$

$Z00z = w + A00invUbd \times \{Kbd^{-1} \times [Ltint \times (Vbd' \times w)]\}$

$sumtr = \mathbf{sum}(z' \times Z00z)$

//Calculate each column's $z_i^\dagger \times Z00 \times z_i$

$trZ00 = sumtr/m$

**return** trZ00